\documentclass[twocolumn,showpacs,amsmath,amssymb,prb]{revtex4}

\pdfoutput=1

\usepackage{amsmath,amsfonts,amssymb,bm}
\usepackage{dcolumn}
\usepackage[final]{graphicx}
\usepackage{comment}
\usepackage{bm}

\begin{document}

\bibliographystyle{apsrev}	

\title{Cavity quantum electrodynamics with semiconductor quantum dots: The role of phonon-assisted cavity feeding}

\author{Ulrich Hohenester}%
  \email{ulrich.hohenester@uni-graz.at}%
\affiliation{Institut f\"ur Physik,
  Karl--Franzens--Universit\"at Graz, Universit\"atsplatz 5,
  8010 Graz, Austria}



\date{December 28, 2009}

\begin{abstract}

For a semiconductor quantum dot strongly coupled to a microcavity, we theoretically investigate phonon-assisted transitions from the exciton to a cavity photon, where the energy mismatch is compensated by phonon emission or absorption. By means of a Schrieffer-Wolff transformation we derive an effective Hamiltonian, which describes the combined effect of exciton-cavity and exciton-phonon coupling, and compute the scattering rates within a Fermi golden rule approach. The results of this approach are compared with those of a recently reported description scheme based on the independent Boson model [U. Hohenester et al., Phys. Rev. B \textbf{80}, 201311(R) (2009)], and a numerical density matrix approach. All description schemes are shown to give very similar results. We present results for the spontaneous emission lifetime of a quantum dot initially populated with a single exciton or biexciton, and for the spectral properties of an optically driven dot-cavity system operating in the strong coupling regime. Our results demonstrate that phonon-assisted feeding plays a dominant role for strongly coupled dot-cavity systems, when the detuning is of the order of a few millielectron volts.

\end{abstract}

\maketitle


\section{Introduction}

Cavity quantum electrodynamics (QED) investigates hybrid systems consisting of a quantum emitter coupled to a quantized cavity mode. It has been first implemented with atoms in high-finesse optical resonators.~\cite{turchette:95,raimond:01,mabuchi:02} Semiconductor systems allow for the design of monolithic structures.~\cite{khitrova:06} A particularly promising approach is based on semiconductor quantum dots embedded in microcavities, such as micropillars,~\cite{gerard:98,michler:00,reithmaier:04,press:07} microdisks,~\cite{gayral:99} or photonic crystal nanocavities.~\cite{hennessy:07,kaniber:08} With theses systems many cavity QED phenomena have been observed, including Purcell enhancement,~\cite{purcell:46,gerard:98} photon bunching and antibunching,~\cite{press:07} single and entangled photon sources,~\cite{michler:00,stace:03,stevenson:06,akopian:06} and strong light matter coupling.~\cite{reithmaier:04,yoshie:04,hennessy:07} Possible applications range from ultralow threshold nanolasers,~\cite{nomura:06} over efficient single and entangled photon sources, to various application in quantum information science.~\cite{mabuchi:02}

In contrast to atoms, semiconductor quantum dots, sometimes referred to as {\em artificial atoms},\/ are in intimate contact with their solid state environment. This leads to enhanced scattering and dephasing contributions. Indeed, cavity-QED experiments with semiconductor quantum dots have revealed a significant feeding channel for the cavity mode,~\cite{hennessy:07,press:07,laucht:09b} even for largely detuned dot-cavity systems, which has been attributed to the influence of the environment. However, the microscopic origin of such feeding has been discussed controversially in the literature. Shake-up processes in charged quantum dots~\cite{kaniber:08} and quasi-continuum excitonic transitions~\cite{winger:09} have been suggested as possible processes for populating the cavity mode, even in presence of large detuning of several tens of meV. For smaller detunings, of the order of a few meV, phonon processes are expected to play an important role. It is now well established that phonon dephasing~\cite{borri:01,krummheuer:02} and relaxation~\cite{hameau:99,verzelen:02,zibik:09} govern the coherent optical response of quantum dots. As for cavity-QED, theoretical work has suggested that such pure phonon dephasing might be responsible for cavity feeding at detunings of the order of the polariton linewidths.~\cite{naesby:08,milde:08,auffeves:09,hughes:09,tarel:09} Experimental work has confirmed that for such small detunings dephasing plays a crucial role.~\cite{suffcynski:09,ates:09} 

In a recent paper,~\cite{hohenester.prb:09} it has been demonstrated theoretically and experimentally that, besides phonon dephasing, there exists another phonon-mediated feeding channel for the cavity that works efficiently over a broad detuning range of several meV. Similar results have been independently obtained by other authors.~\cite{ota:09,kaer:09} The cavity feeding relies on a combined effect of exciton-cavity and exciton-phonon couplings, and accounts for a process where the exciton decays into a cavity photon and the energy mismatch is compensated by the emission or absorption of a phonon. 

In this paper, we present the detailed analysis of the theoretical framework given in Ref.~\onlinecite{hohenester.prb:09}, and provide a simple description scheme for phonon-assisted cavity feeding. More specifically, we bring, by means of a unitary Schrieffer-Wolff transformation, the electron-phonon hamiltonian to a form that precisely describes the combined effect of exciton-cavity and exciton-phonon couplings. We compare our results with those of the previously used independent Boson model and of a numerical density matrix approach, and find almost identical results for all approaches. We also present results for the decay of excitons and biexcitons in presence a phonon-assisted cavity feeding, and analyze the impact of such feeding on the spectral properties of a driven dot-cavity system.

We have organized our paper as follows. In Sec.~\ref{sec:theory} we present the theoretical description scheme for phonon-assisted cavity feeding in semiconductor quantum dots. We start with the Schrieffer-Wolff transformation, where the cavity and phonon couplings are treated as perturbations, which allows us to compute the feeding rates by means of a simple Fermi-golden-rule approach. Due to the simple structure of the independent Boson hamiltonian, one can also describe the phonon dynamics exactly and only treat the exciton-cavity coupling as a perturbation. We finally introduce a density matrix approach, that allows for simulations without employing any approximation. Some details of this approach are presented in the appendix. In Sec.~\ref{sec:results} we present results for these different description schemes. We discuss the feeding rates and the resulting spontaneous emission lifetimes for different detunings, temperatures, and cavity quality factors. We also investigate the spectra for a driven dot-cavity system, and analyze the detuning dependence of the polariton intensities. Our results demonstrate that phonon-assisted feeding plays a dominant role for strongly coupled dot-cavity systems when the detuning is of the order of a few meV. Finally, in Sec.~\ref{sec:summary} we summarize and draw some conclusions.


\section{Theory}\label{sec:theory}

\subsection{Model}

\subsubsection{Hamiltonian}

In the following we consider a quantum dot embedded in a microcavity, which interacts with phonons.~\cite{wilson-rae:02,hohenester.prb:09} We will show that, due to the coupling of the quantum dot state with the phonons, excitonic states can decay into cavity photons even for relatively large exciton-cavity detunings of a few meV. We start with a description of the excitonic states in terms of a generic two level system, where $e$ denotes the excited and $g$ the ground state. Although we shall be primarily concerned with the exciton to groundstate decay, our description scheme also applies to the biexciton to exciton decay. With the quantum dot operators $\sigma_{gg}=|g\rangle\langle g|$ and $\sigma_{ee}=|e\rangle\langle e|$, and $a$ and $b_k$ the bosonic annihilation operators for the cavity mode and the $k$th phonon mode (with energies $\omega_{\rm cav}$ and $\omega_k$), the Hamiltonian for the uncoupled system reads
\begin{equation}\label{eq:ham0}
  H_0=\bigl(E_g\sigma_{gg}+E_e\sigma_{ee}\bigr)+\omega_{\rm cav}\,a^\dagger a+\sum_k
  \omega_k\,b_k^\dagger b_k^{\phantom\dagger}\,.
\end{equation}
Here $E_g$ and $E_e$ are the energies of the ground and excited quantum dot state. We set $\hbar=1$ throughout this paper. The exciton-cavity coupling in the rotating-wave approximation~\cite{walls:95,scully:97} is of the form
\begin{equation}\label{eq:hamg}
  H_g=g\left(\sigma_{eg}a+\sigma_{ge}a^\dagger\right)\,,
\end{equation}
where $\sigma_{ge}=|g\rangle\langle e|$ and $\sigma_{eg}=|e\rangle\langle g|$ are the lowering and raising operators for the quantum dot, and $g$ is the coupling constant, which depends on the exciton dipole moment and the field distribution of the cavity mode.~\cite{andreani:99,walls:95} Through $H_g$ exciton and cavity become coupled, and new polariton modes are formed which have partial exciton and partial cavity character.~\cite{walls:95,reithmaier:04,hennessy:07} In the strong coupling regime, the coupling is stronger than the quantum dot and cavity losses, and excitation is coherently transferred between the dot and the cavity. In contrast, in the weak coupling regime losses dominate over the coherent coupling, and the excited state decays mono-exponentially. Quite generally, the calculation of the phonon-assisted scattering rates, to be discussed below, works in both the strong and weak coupling regime, and we thus will not be specific about this point until Sec.~\ref{sec:master} where we introduce our master-equation approach.

The coupling between the quantum dot states and the phonons is described by the generic electron-phonon Hamiltonian~\cite{mahan:81,wilson-rae:02,hohenester.jpb:07}
\begin{equation}\label{eq:hamib}
  H_{\rm ep}=\sigma_{gg}B_{gg}+\sigma_{ee}B_{ee}\,,
\end{equation}
where we have introduced the {\em bath operators}\/~\cite{breuer:02}
\begin{equation}\label{eq:opbath}
  B_{ij}=\sum_k\lambda_{ij,k}\left(b_k^{\phantom\dagger}+b_k^\dagger \right)\,.
\end{equation}
Here, $\lambda_{ij,k}$ denotes the electron phonon matrix elements. In case of the exciton to groundstate decay $B_{gg}$ is zero, since only the exciton interacts with the lattice degrees of freedom, whereas for the biexciton to exciton decay both $B_{gg}$ and $B_{ee}$ are non-zero. The diagonal bath operators $B_{gg}$ and $B_{ee}$ describe a coupling where the lattice becomes distorted, but no transitions between quantum dot states are induced. The corresponding model is usually denoted as the {\em independent Boson model}.~\cite{mahan:81,krummheuer:02,wilson-rae:02,hohenester.jpb:07} In Eq.~\eqref{eq:opbath} we have defined the bath operators in a slightly more general way, such that also transitions between different quantum dot states can be described, for reasons that will become clear further below.

\subsubsection{Scattering rate}

To compute the phonon-assisted scattering rates between the quantum dot and the cavity, we employ a master equation approach. Let 
\begin{equation}\label{eq:hint}
  H=\bar H_0+V
\end{equation}
be the Hamiltonian composed of a part $\bar H_0$, which can be treated exactly, and a perturbation part $V$. In the most simple approach, $\bar H_0$ is associated with the Hamiltonian of Eq.~\eqref{eq:ham0}, and $V$ with the exciton-cavity and exciton-phonon couplings of Eqs.~\eqref{eq:hamg} and \eqref{eq:hamib}. However, as we will show in Sec.~\ref{sec:ib}, due to the simple structure of the independent Boson hamiltonian of Eq.~\eqref{eq:hamib} we can even describe the full dynamics of $H_0+H_{\rm ep}$ exactly, and only consider $H_g$ as a perturbation. 

The time evolution of the composite quantum-dot cavity system, described by the density matrix $\rho$, and the phonon part $\rho_{\rm ph}$, can be approximately described by the master equation in Born approximation \cite{walls:95,breuer:02}
\begin{equation}\label{eq:born}
  \frac{d\rho(t)}{d t}
  \approx-\int_{t_0}^t\,\mbox{tr}_{\rm ph}\bigl\lgroup
  [V(t),[V(\tau),\rho(t)\otimes \rho_{\rm ph}]\,]
  \bigr\rgroup d\tau\,,
\end{equation}
where $V(t)$ is the perturbation in the interaction representation of $\bar H_0$. In Eq.~\eqref{eq:born} we have assumed a factorizable density matrix~\cite{breuer:02} at the initial time $t_0$. In the following we consider transitions between the initial and final states $i$ and $f$ of the quantum-dot cavity system, which, e.g.,  for the phonon-assisted decay read $|e;0\rangle$ and $|g;1\rangle$. Here, the first entry corresponds to the quantum dot state and the second one to the number of photons in the cavity. Upon explicit evaluation of the double-commutator in Eq.~\eqref{eq:born} we encounter contributions where both $V$ operators act from either the left or right hand side on $\rho$, which can be associated with generalized out-scatterings and dephasing.~\cite{breuer:02,hohenester.review:06} The other contributions can be associated with generalized in-scatterings. More specifically, we find for the increase of population in the final state
\begin{equation}\label{eq:dotrhoff}
  \dot\rho_{f\!f}=
  \int_{-\infty}^0\mbox{tr}_{\rm ph}\bigl\lgroup
  \langle f|V(0)|i\rangle\rho_{ii}\rho_{\rm ph}
  \langle i|V(\tau)|f\rangle+\mbox{c.c.}\bigr\rgroup d\tau\,,
\end{equation}
where $\mbox{c.c.}$ denotes the complex conjugate of the preceding term. We have used the adiabatic approximation of letting the lower integration limit approach minus infinity and introducing a small damping term inside the integral (not shown), which allows us to perform the time integration prior to the phonon trace.~\cite{hohenester.review:06} From Eq.~\eqref{eq:dotrhoff} we can infer the scattering rate 
\begin{equation}\label{eq:scattrate}
  \Gamma_{if}=2\,\mbox{Re}\int_{-\infty}^0\mbox{tr}_{\rm ph}\bigl\lgroup
  \rho_{\rm ph}
  \langle i|V(\tau)|f\rangle\langle f|V(0)|i\rangle\bigr\rgroup d\tau
\end{equation}
for a transition from the initial state $i$ to the final state $f$. To arrive at  our final expression, we have exploited the cyclic permutation of operators under the trace. The generalized Fermi golden rule expression of Eq.~\eqref{eq:scattrate} will serve us as the starting point for the calculation of the phonon-assisted scattering rates.

\subsection{Schrieffer--Wolff transformation}\label{sec:schrieffer}

The decay of an exciton to the cavity under off-resonant conditions requires both the coupling between states $e$ and $g$, by means of the exciton-cavity hamiltonian $H_g$, and the emission or absorption of a phonon, in order to compensate for the energy mismatch. The scattering rate will thus depend on both cavity and phonon couplings $g$ and $\lambda_k$. A convenient way to account for the combined cavity-phonon character of the scattering is to remove, in lowest perturbation order, the exciton-cavity coupling by means of a Schrieffer--Wolff transformation.~\cite{schrieffer:66,grond:08} Let us consider first the Hamiltonian $H=H_0+H_g$ composed of the free part and the exciton-cavity coupling. By means of a unitary transformation~\cite{grond:08}
\begin{equation}\label{eq:unitary}
  \tilde H=e^sHe^{-s}=H+[s,H]+\frac 12[s,[s,H]]+\dots
\end{equation}
we can remove the exciton-cavity coupling to the lowest order of $g$, by choosing $s$ such that $[s,H_0]+H_g=0$. This is achieved through the operator
\begin{equation}
  s=\frac g\Delta\left(\sigma_{eg}a-\sigma_{ge}a^\dagger\right)\,,
\end{equation}
where $\Delta=E_e-E_g-\omega_{\rm cav}$ is the detuning between the quantum dot transition and the cavity. Then, $\tilde H=H_0+\mathcal{O}(g^2)$ describes the uncoupled exciton-cavity system, and the second-order corrections can be identified as Lamb and Stark shifts~\cite{grond:08} which will be neglected below.

\subsubsection{Diagonal electron-phonon coupling}

Let us investigate how the unitary transformation of Eq.~\eqref{eq:unitary} affects the electron-phonon coupling. After some simple calculation we arrive at
\begin{equation}\label{eq:phonariton}
  \tilde H_{\rm ep}\approx H_{\rm ep}-\frac g\Delta
  \left(\sigma_{eg}a-\sigma_{ge}a^\dagger\right)
  \bigl(B_{ee}-B_{gg}\bigr)\,
\end{equation}
for the lowest order correction of $\tilde H_{\rm ep}$, which is the expression we are seeking for. The second term on the right-hand side describes a combined interaction of $H_g$, which couples the quantum dot states $e$ and $g$ via emission or absorption of a cavity photon, and $H_{\rm ep}$, which can compensate for the energy mismatch $\Delta$ via emission or absorption of a phonon. In what follows, we assign the perturbation $V$ to the Hamiltonian given by Eq.~\eqref{eq:phonariton}, and consider an interaction representation according to $H_0$. Upon insertion of the Hamiltonian of Eq.~\eqref{eq:phonariton} into the scattering rate of Eq.~\eqref{eq:scattrate}, we obtain for the transition from the initial state $|e;0\rangle$ to the final state $|g;1\rangle$ the rate
\begin{equation}
  \Gamma_{eg}=2\,\mbox{Re}\left\lgroup\left(\frac g\Delta\right)^2\int_{-\infty}^0
  e^{i(\Delta-i0^+)\tau}\left<\delta B(\tau)\delta B(0)\right>d\tau\right\rgroup\,,
\end{equation}
with $0^+$ being a small positive quantity and $\delta B=B_{ee}-B_{gg}$ the bath fluctuation operator. We have used $\left<\cdot\right>=\mbox{tr}_{\rm ph}(\rho_{\rm ph}\cdot)$. In thermal equilibrium only the phonon occupations $\langle b_k^\dagger b_k^{\phantom\dagger}\rangle=\bar n_k$ are non-zero ($\bar n_k$ is the Bose-Einstein distribution at temperature $T$), and we can evaluate the phonon correlation function analytically.~\cite{breuer:02} We then get 
\begin{eqnarray}\label{eq:feedingrate}
  &&\Gamma_{eg}=\left(\frac g\Delta\right)^2
  2\pi\sum_k\left|\lambda_{ee,k}-\lambda_{gg,k}\right|^2\nonumber\\ 
  &&\qquad\times
  \bigl\lgroup\bar n_k\delta(\Delta+\omega_k)+(\bar n_n+1)\delta(\Delta-\omega_k)\bigr\rgroup
  \quad
\end{eqnarray}
for the phonon-assisted feeding rate of the cavity through decay of the exciton state. Eq.~\eqref{eq:feedingrate} is our final expression. It accounts for the phonon-assisted exciton decay to the cavity, where the phonons are assumed to be in thermal equilibrium.

If the exciton decays within a cavity populated with $n$ photons, the scattering rate has to be multiplied by $n$ (because of the cavity field operators acting on the Fock state with $n$ photons). We can also derive along the same lines the scattering rate $\Gamma_{ge}$ for the reversed process, where the photon is removed from the cavity and an exciton is created. The resulting expression is identical to Eq.~\eqref{eq:feedingrate} except that $\Delta$ has to be replaced by $-\Delta$. This backscattering, however, is usually inefficient due to the strong losses of the cavity.

\subsubsection{Off-diagonal electron-phonon coupling}

Our above derivation can be also applied to the case where more than two quantum dot states are involved. Indeed, it has been demonstrated that in the cascade multi-exciton decay of an optically excited quantum dot the emission of cavity photons plays an important role.~\cite{kaniber:08,winger:09} Since there is practically always an energy mismatch between the transition energies of the quantum dot states and the cavity photon, phonons are expected to play a decisive role also in this cascade. As a simple model for a phonon-assisted cascade decay, we consider again two states $e$ and $g$ which are electromagnetically coupled through $H_g$. However, for the phonon coupling we assume that the lower state $g$ is coupled through phonon emission or absorption to some other state $g'$ viz.
\begin{equation}
  H_{\rm ep}'=\bigl(\sigma_{gg'}+\sigma_{g'g}\bigr)B_{gg'}\,.
\end{equation}
The additional state $g'$ might be even optically inactive, corresponding, e.g., to the excitation of one hole or electron out of the groundstate of a multi-exciton state. It turns out that for this off-diagonal phonon Hamiltonian the scattering rate can be computed in a completely similar fashion, and we arrive at
\begin{eqnarray}\label{eq:feedingrate2}
  &&\Gamma_{eg'}=\left(\frac g\Delta\right)^2
  2\pi\sum_k\left|\lambda_{gg',k}\right|^2\nonumber\\ 
  &&\qquad\times
  \bigl\lgroup\bar n_k\delta(\Delta'+\omega_k)+(\bar n_n+1)\delta(\Delta'-\omega_k)\bigr\rgroup\,,
  \quad
\end{eqnarray}
where $\Delta'=E_e-E_{g'}-\omega_{\rm cav}$ is the detuning with respect to state $g'$. Equation~\eqref{eq:feedingrate2} accounts for an indirect transition, where the system decays via the optical channel to $g$, and becomes promoted through phonon emission or absorption to the final state $g'$. As a result, multi-exciton states can easily decay even in absence of strict energy conservation for the exciton-cavity system.

\subsection{Independent Boson model}\label{sec:ib}

Due to the particular structure of the independent boson Hamiltonian \eqref{eq:hamib} it is possible to treat the electron-phonon interaction without further approximation. In this approach, we introduce an interaction representation according to $\bar H_0=H_0+H_{\rm ep}$ and treat the exciton-photon coupling $H_g$ as the perturbation. 

The key to the success of the independent Boson model is provided by the phonon displacement operator,~\cite{walls:95,scully:97,hohenester.jpb:07} which, for a single phonon mode, reads $e^s$ with $s=\lambda/\omega(b-b^\dagger)$. Through the unitary transformation $e^s be^{-s}=b+(\lambda/\omega)$ the phonon field becomes displaced by a constant value. Then,
\begin{eqnarray}
  e^s\Bigl(\omega\,b^\dagger b\Bigr)e^{-s}&=&
  \omega\left(b^\dagger+\frac \lambda\omega\right)
  \left(b+\frac\lambda\omega\right)\nonumber\\
  &=&\omega\,b^\dagger b+\lambda\Bigl(b+b^\dagger\Bigr)+\frac{\lambda^2}{\omega}
\end{eqnarray}
gives the free-phonon hamiltonian for the displaced phonon fields. Let us first assume that only the upper state interacts with the phonons, i.e., $H_{\rm ep}=\sigma_{ee}B_{ee}$ (the case where phonon coupling is also present in the lower state, such as for the biexciton to exciton decay, will be addressed below). The transformation $e^{\sigma_{ee}s}$ with
\begin{equation}\label{eq:s.ib}
  s=\sum_k\frac{\lambda_{ee,k}}{\omega_k}\left(b_k^{\phantom\dagger}-b_k^\dagger\right)
\end{equation}
then allows to generate the independent boson hamiltonian through~\cite{hohenester.jpb:07} $H_0+H_{\rm ep}\approx e^{\sigma_{ee}s}H_0e^{-\sigma_{ee}s}$, apart from an energy renormalization $-\sigma_{ee}\sum_k\lambda_{ee,k}^2/\omega_k$ of the quantum dot state, which is usually referred to as the polaron shift and which can be easily absorbed into the definition of the dot energies.~\cite{mahan:81,krummheuer:02,hohenester.jpb:07} This transformation is known as the {\em polaron transformation},\/ as it displaces, depending on the state of the quantum dot, the phonon fields to the new equilibrium positions. For the exciton decay, the groundstate is associated with an undistorted lattice, and only for the exciton state the lattice becomes distorted, due to the formation of a polaron. Then, we obtain from Eq.~\eqref{eq:scattrate} the scattering rate
\begin{eqnarray}\label{eq:scattrate.ib0}
  &&\Gamma_{eg}=2g^2\,\mbox{Re}\int_{-\infty}^0\mbox{tr}_{\rm ph}
  \nonumber\\ &&\quad\times
  \bigl\lgroup\rho_{\rm ph}
  \bigl<g;1\bigr|e^{iH_0\tau}\bigl|g;1\bigr\rangle\bigl\langle e;0\bigr|
  e^{s}e^{-iH_0\tau}e^{-s}\bigl|e;0\bigr>\bigr\rgroup d\tau\,, \nonumber\\
\end{eqnarray}
where we have used $e^{\sigma_{ee}s}=\sigma_{gg}+\sigma_{ee}e^s$. We next evaluate all matrix elements in Eq.~\eqref{eq:scattrate.ib0} to transform the integrand to $e^{-i\Delta\tau}\left<e^{s(\tau)}e^{-s}\right>$, where $s(\tau)$ is the expression of Eq.~\eqref{eq:s.ib} in the interaction representation of $H_0$ and the brackets denote the expectation value over the phonon density matrix. In thermal equilibrium the expectation value $\left<e^{-s(\tau)}e^{s}\right>$ can be evaluated analytically~\cite{mahan:81,breuer:02,hohenester.jpb:07,remark.correlation}
\begin{eqnarray}\label{eq:correlation}
  &&C(\tau)=\nonumber\\ &&\quad
  \exp\bigl\lgroup-\sum_k\left(\frac{\lambda_{ee,k}}{\omega_k}\right)^2
  \left[(\bar n_k+1)e^{-i\omega_k\tau}+\bar n_k 
  e^{i\omega_k\tau}\right]\bigr\rgroup\,.\nonumber\\
\end{eqnarray}
This correlation function also determines pure phonon dephasing and the spectral lineshape in optically excited semiconductor quantum dots.~\cite{borri:01,krummheuer:02} The phonon-assisted scattering rate within the independent Boson model then becomes~\cite{hohenester.prb:09}
\begin{equation}\label{eq:scattrate.ib}
  \Gamma_{eg}=2g^2\,\mbox{Re}\int_{-\infty}^0 e^{-i\Delta\tau}C(-\tau)\,d\tau\,.
\end{equation}
One can easily show that, by expanding the integrand in Eq.~\eqref{eq:correlation} to lowest order, we precisely recover the result of Eq.~\eqref{eq:feedingrate} for the Schrieffer-Wolff transformation. Finally, when phonon interactions are present in both the excited and the ground state, we have to replace in Eq.~\eqref{eq:correlation} the phonon matrix elements simply by $\lambda_{ee,k}-\lambda_{gg,k}$. This is due to the fact that only the \textit{difference} in phonon equilibrium positions affects the phonon dephasing and, in turn, the phonon assisted scattering rates.

\subsection{Master equation approach}\label{sec:master}

\begin{figure}
\includegraphics[width=\columnwidth]{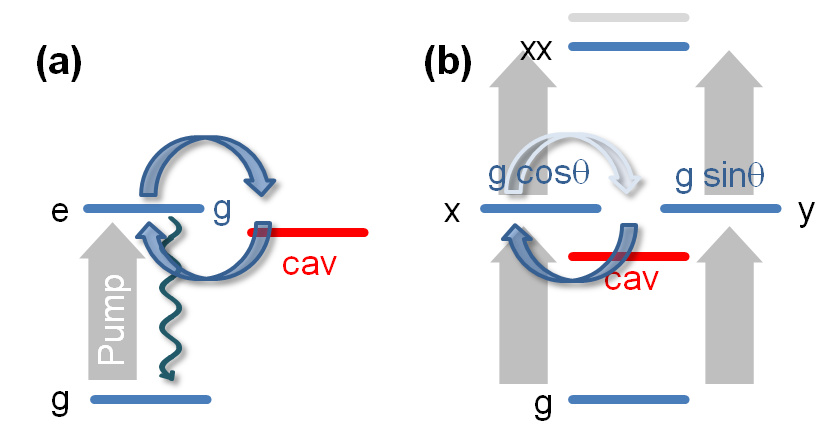}
\caption{(Color online) Level schemes used in our simulations. (a) Ground and excited quantum dot states $e$, $g$, and cavity. The exciton state is coupled to the cavity, with the cavity coupling $g$. The exciton and cavity decay with rates $\gamma_{\rm rad}$ and $\kappa$. In case of laser pumping, we also consider transitions from $g$ to $e$. (b) Biexciton level scheme, consisting of the dot groundstate $g$, the exciton states $x$ and $y$, with orthogonal polarizations, and the biexciton state $X\!X$. In our simulations we use a basis of the bare quantum dot states, and these states together with up to two photons in the cavity. When the cavity and the dot are not perfectly aligned, both exciton transitions can couple to the cavity. We assume that the $X\!X\to x$ and $x\to g$ transitions couple with strength $g\cos\theta$, and the $X\!X\to y$ and $y\to g$ transitions with strength $g\sin\theta$ to the cavity. Here $\theta$ is an angle that determines the degree of mixing. Pumping brings the system from the ground state via the exciton states $x$ and $y$ to the biexciton state $X\!X$. }\label{fig:scheme}
\end{figure}

With the phonon assisted scattering rates, obtained within either the Schrieffer-Wolff or independent Boson model approach, we can describe the dynamics of the coupled dot-cavity system by means of a {\em master equation approach}.~\cite{walls:95} In this work we will consider the level schemes depicted in Fig.~\ref{fig:scheme}, which consist of a different number of composite dot-cavity states. We describe the coherent dynamics by the Hamiltonian $H$, which accounts for the single-particle energies and the dot-cavity couplings. The incoherent dynamics, which accounts for radiative dot and cavity losses, phonon-assisted cavity feeding, and dot pumping in presence of an external laser pulse, is described by the Lindblad operators~\cite{walls:95,breuer:02} $L_\mu$, where $\mu$ labels the different scattering channels. The time evolution of the density matrix is then determined by a master equation of Lindblad form~\cite{walls:95,breuer:02}
\begin{equation}\label{eq:master}
  \dot\rho=-i[H,\rho]-\sum_\mu\left(\frac 12\left\{L_\mu^\dagger 
  L_\mu^{\phantom\dagger},\rho\right\}-L_\mu^{\phantom\dagger}\rho L_\mu^\dagger\right)\,.
\end{equation} 
Depending on the relative importance of the dot-cavity coupling $g$ and the various scattering losses, the solutions of this master equation exhibit strong or weak coupling effects. The master equation of Eq.~\eqref{eq:master} can be rewritten in the form $\dot\rho=-i\mathsf{L}\rho$, where $\mathsf{L}$ is the Liouvillian of the dot-cavity system. If we interpret $\rho$ as a column vector, with matrix elements $\rho_{ij}$ ordered in some unique fashion, $\mathsf{L}$ becomes a matrix, whose elements can be directly read off from Eq.~\eqref{eq:master}. A convenient way to solve the master equation is to seek for the right and left eigenvectors $\mathsf{X}$ and $\tilde{\mathsf{X}}$, which are defined through~\cite{press:02}
\begin{equation}\label{eq:mastereig}
  \mathsf{L}\mathsf{X}=\mathsf{X}\mathsf{\Lambda}\,,\quad
  \tilde{\mathsf{X}}\mathsf{L}=\mathsf{\Lambda}\tilde{\mathsf{X}}\,,
\end{equation}
where $\mathsf{\Lambda}$ is a diagonal matrix containing the eigenvalues. The eigenvectors form two biorthogonal sets with $\mathsf{X}\tilde{\mathsf{X}}= \tilde{\mathsf{X}}\mathsf{X}=\openone$. The master equation, subject to the initial condition $\rho(0)=\rho_0$, can then be easily solved through
\begin{equation}\label{eq:master.time}
  \rho(t)=\left(\mathsf{X}e^{-i\mathsf{\Lambda}t}\tilde{\mathsf{X}}\right)\rho_0\,,
\end{equation}
with the expression in parenthesis being the time evolution operator for the density matrix. Eq.~\eqref{eq:master.time} is extremely useful for computational purposes, since it allows for the full solution of the master equation by simply diagonalizing once the $\mathsf{L}$ matrix.

From the solutions of the eigenvalue problem of Eq.~\eqref{eq:mastereig} we can also obtain analytic expressions for the spectrum of the coupled dot-cavity system. We start from the Wiener-Khinchin theorem,~\cite{walls:95,scully:97} which, in case of continuous dot pumping, reads
\begin{equation}\label{eq:wienerkhinchin}
  S(\omega)\propto\mbox{Re}\int_0^\infty e^{i\omega\tau}
  \bigl< a^\dagger(0)a(\tau)\bigr>_{\rm s.s.}\,d\tau\,,
\end{equation}
where the brackets denote the expectation value at late times when the system has reached its steady state. Eq.~\eqref{eq:wienerkhinchin} relates the spectrum to the two-time correlation function of the cavity fields, which can be computed by means of the {\em quantum regression theorem}.~\cite{gardiner:04} Let $A$ be the matrix of expectation values for the cavity field operator $a$ between the composite dot-cavity states of our master equation approach. The quantum regression theorem then allows us to express the correlation function as
\begin{equation}\label{eq:regression}
  \bigl< a^\dagger(0)a(\tau)\bigr>_{\rm s.s.}=
  \mbox{tr}\bigl\lgroup A_{ji}
  \left(\mathsf{X}e^{-i\mathsf{\Lambda}\tau}\tilde{\mathsf{X}}\right)_{ij,kl}
  \left(\rho_{ss}A^\dagger\right)_{kl}\bigr\rgroup\,,
\end{equation}
where we have implicitly assumed summation over repeated indices. $\rho_{ss}$ denotes the density matrix in the steady state, which can be computed as the long time limit of Eq.~\eqref{eq:master.time}.  Equation~\eqref{eq:regression} has to be interpreted as follows.~\cite{gardiner:04} First, the operator at the earlier time, $a^\dagger$, acts from the right-hand side on $\rho_{ss}$, thereby creating a fluctuation in the system. Next, this fluctuation propagates in time, which is described by the time evolution operator of Eq.~\eqref{eq:master.time}. Finally, we determine the expectation value for the operator at the later time, $a$, with the propagated fluctuation operator. Again, the knowledge of the eigenvectors and eigenvalues allows us to easily evaluate the two-time correlation function. Upon insertion of the correlation function into the Wiener-Khinchin expression for the spectrum, Eq.~\eqref{eq:wienerkhinchin}, we can compute the spectrum of the driven dot-cavity system according to
\begin{equation}\label{eq:spectrum}
  S(\omega)\propto \text{Im}\,\mbox{tr}\bigl\lgroup A_{ji}
  \left(\mathsf{X}({\mathsf{\Lambda}-\omega})^{-1}\tilde{\mathsf{X}}\right)_{ij,kl}
  \left(\rho_{ss}A^\dagger\right)_{kl}
  \bigr\rgroup\,.
\end{equation}
In a very similar fashion we could also compute photon correlation functions.~\cite{scully:97,laussy:08}

\subsection{Density matrix approach}\label{sec:densitymatrix}

We finally present an approach suited for the numerical solution of the coupled dot-cavity dynamics in presence of phonon interactions, without employing a perturbation approach. We consider the situation where the system is initially in the excited dot state, and subsequently decays through radiative decay or coupling to the cavity photon, which then leaks out of the cavity. In absence of dot pumping, we are left with a generic two-level system consisting of the excited quantum dot state together with the dot groundstate and one photon in the cavity. We can describe the system in terms of an effective Hamiltonian
\begin{equation}\label{eq:heff}
  H_{\rm eff}=\left(\Delta-i\frac\gamma 2\right)\sigma_{ee}-i\frac\kappa 2\sigma_{gg}
  +g\left(\sigma_{eg}+\sigma_{ge}\right)\,,
\end{equation}
where $\gamma$ is the radiative decay rate of the excited quantum dot state and $\kappa$ the leakage rate of the cavity. Note that, in contrast to the above analysis, here $g$ denotes the state composed of the dot groundstate \textit{and} the cavity photon. It is important to realize that any scattering, described by the loss terms proportional to $\gamma$ and $\kappa$, brings the system to a state (dot groundstate without cavity photon) that is not coupled to our two-level system. 

The total Hamiltonian then consists of $H_{\rm eff}$, together with the electron-phonon coupling of Eq.~\eqref{eq:hamib} and the free phonon Hamiltonian. For simplicity, we consider the situation where only the excited state couples to phonons, but our results could be easily generalized if phonon coupling was also present in the dot groundstate.

We next employ a density matrix approach.~\cite{foerstner:03a,foerstner:03b,hohenester.prl:04,hohenester.review:06,hohenester.jpb:07} To this end, we start from the density matrix $\rho_{ij}=\left<\sigma_{ij}\right>$, where the brackets denote $\mbox{tr}(\rho\sigma_{ij})$, and compute its equation of motion from the Heisenberg equation of motion
\begin{equation}
  i\dot\rho_{ij}=\bigl<\sigma_{ij}H_{\rm eff}^\dagger-H_{\rm eff}^{\phantom\dagger}
  \sigma_{ij}\bigr>
  +\bigl<[\sigma_{ij},\sigma_{ee}]B_{ee}\bigr>\,.
\end{equation}
Because of the phonon coupling, the density matrix couples to the {\em phonon-assisted density matrix}~\cite{rossi:02} $\left<\sigma_{ij}b_k\right>$. If we compute next the equation of motion for the phonon assisted density matrix, we find that it couples to a density matrix with two phonon operators, whose time evolution, in turn, involves density matrices with three phonon operators. This reflects the fact that the interacting dot-phonon system constitutes a highly non-trivial many-body problem, with an infinite hierarchy of equations of motion for the density matrices. To truncate this infinite hierarchy, it is convenient to employ a {\em cumulant expansion},\cite{breuer:02,foerstner:03a,hohenester.jpb:07} where one introduces correlation functions by subtracting from the higher-order density matrices contributions of lower-order density matrices. For instance, we obtain for the phonon assisted density matrix
\begin{equation}
  \bigl<\sigma_{ij}b_k\bigr>=\bigl<\sigma_{ij}\bigr>\bigl< b_k\bigr>+
  \bigl<\!\bigl<\sigma_{ij}b_k\bigr>\!\bigr>\,,
\end{equation}
where $\left<\!\left< \cdot\right>\!\right>$ denotes the correlation function. On general physical grounds, we expect that higher-order phonon-assisted correlation functions will play no significant role in the dynamics of $\rho_{ij}$.

\begin{table}[t]
\caption{Correlation functions used in our density matrix approach. We consider all correlation functions which contain less than three phonon operators. The equations of motion for the different correlation functions are given in Appendix~\ref{sec:cumulant}.
}\label{table:cumulant}
\begin{ruledtabular}
\begin{tabular}{lll}
Correlation function & Symbol & Expression  \\
\colrule
Density matrix & $\rho_{ij}$ & $\langle\sigma_{ij}\rangle$ \\
Phonon amplitude & $s_k$ & $\langle b_k^{\phantom\dagger}\rangle$ \\
Phonon-assisted density matrix & $s_{ij,k}$ &
  $\langle\!\langle \sigma_{ij}b_{k}^{\phantom\dagger}\rangle\!\rangle$ \\
Two-phonon amplitude & $s_{kk'}$ & 
  $\langle\!\langle b_k^{\phantom\dagger}b_{k'}^{\phantom\dagger}\rangle\!\rangle$ \\
Phonon occupation & $n_{kk'}$ &
  $\langle\!\langle b_k^\dagger b_{k'}^{\phantom\dagger}\rangle\!\rangle$ \\  
Two-phonon-assisted density matrix & $s_{ij,kk'}$ &
  $\langle\!\langle \sigma_{ij}b_{k}^{\phantom\dagger}b_{k'}^{\phantom\dagger}
  \rangle\!\rangle$ \\
  & $n_{ij,kk'}$ &
  $\langle\!\langle \sigma_{ij}b_{k}^\dagger b_{k'}^{\phantom\dagger}
  \rangle\!\rangle$ \\
\end{tabular}
\end{ruledtabular}
\end{table}

In our density matrix equation approach we first select a set of representative correlation functions. Here we restrict ourselves, in accordance to related work,\cite{foerstner:03a,hohenester.jpb:07} to those correlation functions which contain less than three phonon operators (see Table \ref{table:cumulant} for a complete list). Their equations of motions are obtained by using the Heisenberg equation of motion, performing the cumulant expansion, and keeping only the relevant correlation functions. By neglecting higher-order correlations, we truncate the infinite hierarchy of equations of motions, and obtain a closed set of equations of motion (see Appendix~\ref{sec:cumulant}). This set can be solved through direct numerical integration. Results of these calculations, and a comparison with the perturbative approach, will be given in the next section.


\section{Results}\label{sec:results}

\subsection{Model}

\begin{table}[t]
\caption{Material, dot, and cavity parameters used in our simulations. The phonon parameters are representative for GaAs. For the electron and hole wavefunctions, we assume Gaussians with a full width of half maximum (FWHM) of $L_{\rm lat}$ and $L_z$ along the lateral and growth direction, respectively. Throughout we use an exciton-cavity coupling strength of 150 $\mu$eV and a radiative dot lifetime of 7 ns. The cavity quality factor is $Q=15000$ unless stated differently.
}\label{table:param}
\begin{ruledtabular}
\begin{tabular}{lll}
Parameter & Symbol & Value  \\
\colrule
mass density & $\rho$ & 5.37 g/cm$^{-3}$ \\
sound velocity & $c_\ell$ & 5110 m/s\\
deformation potential for electrons & $D_e$ & $-14.6$ eV \\
deformation potential for holes & $D_h$ & $-4.8$ eV \\
in-plane confinement (FWHM) & $L_{\rm lat}$ &  10 nm\\
confinement in $z$-direction (FWHM) & $L_z$ &  4 nm \\
biexciton binding energy & $\Delta_b$ & 2 meV \\
radiative decay time of dot & $\tau_{\rm rad}$ & 7 ns \\
exciton-cavity coupling & $g$ & 150 $\mu$eV \\
cavity photon energy & $\omega_{\rm cav}$ & 1.3 eV \\
\end{tabular}
\end{ruledtabular}
\end{table}

In our simulations we use a quantum dot and phonon model similar to Ref.~\onlinecite{krummheuer:02}, with Gaussian electron and hole wavefunctions, and phonon parameters representative for GaAs (see Table~\ref{table:param}). Quite generally, we expect that a more realistic description scheme, including effects due to strain, the more complicated valence band structure, or the heterogeneous material composition, might somewhat alter our results. On the other hand, our model has proven successful in reproducing the dominant effects of phonon dephasing,~\cite{krummheuer:02,foerstner:03a, hohenester.jpb:07} and is sufficiently generic to include possible modifications from a microscopic model description by simply adapting the effective parameters (quantum dot form factor, phonon coupling). We only consider deformation potential phonon coupling, which is expected to be the dominant contribution. For the cavity feeding through decay of the exciton, the exciton phonon coupling is of the form~\cite{mahan:81,krummheuer:02}
\begin{equation}\label{eq:epmatrix}
  \lambda_{ee,k}=\sqrt{\frac k{2\rho c_\ell}}
  \int e^{-i\bm k\cdot\bm r}
  \left(D_e|\phi_e(\bm r)|^2-D_h|\phi_h(\bm r)|^2\right)\,d^3r\,,
\end{equation}
where $k$ denotes the wavevector, and $\phi_{e,h}(\bm r)$ are the electron and hole wavefunctions for which we assume Gaussians (see Table~\ref{table:param} for parameters). Piezoelectric coupling, described by performing angular averages over the longitudinal and transverse modes separately,~\cite{vorojtsov:05,hohenester.prb:06} turned out to give only negligible corrections, and was thus neglected. Coupling to LO phonons leads to the formation of a tightly bound polaron.~\cite{hameau:99,verzelen:02} The resulting energy shift can be easily absorbed into the exciton and biexciton energies. The polaron-mediated relaxation channel for excited states, provided by the anharmonic decay of LO phonons, is expected to be of minor importance since we are dealing with the exciton and biexciton ground states (where polarons are stable) and are only interested in energy transfers much smaller than the LO phonon energy.

\subsection{Phonon-assisted cavity feeding}

\begin{figure}
\centerline{\includegraphics[width=0.95\columnwidth]{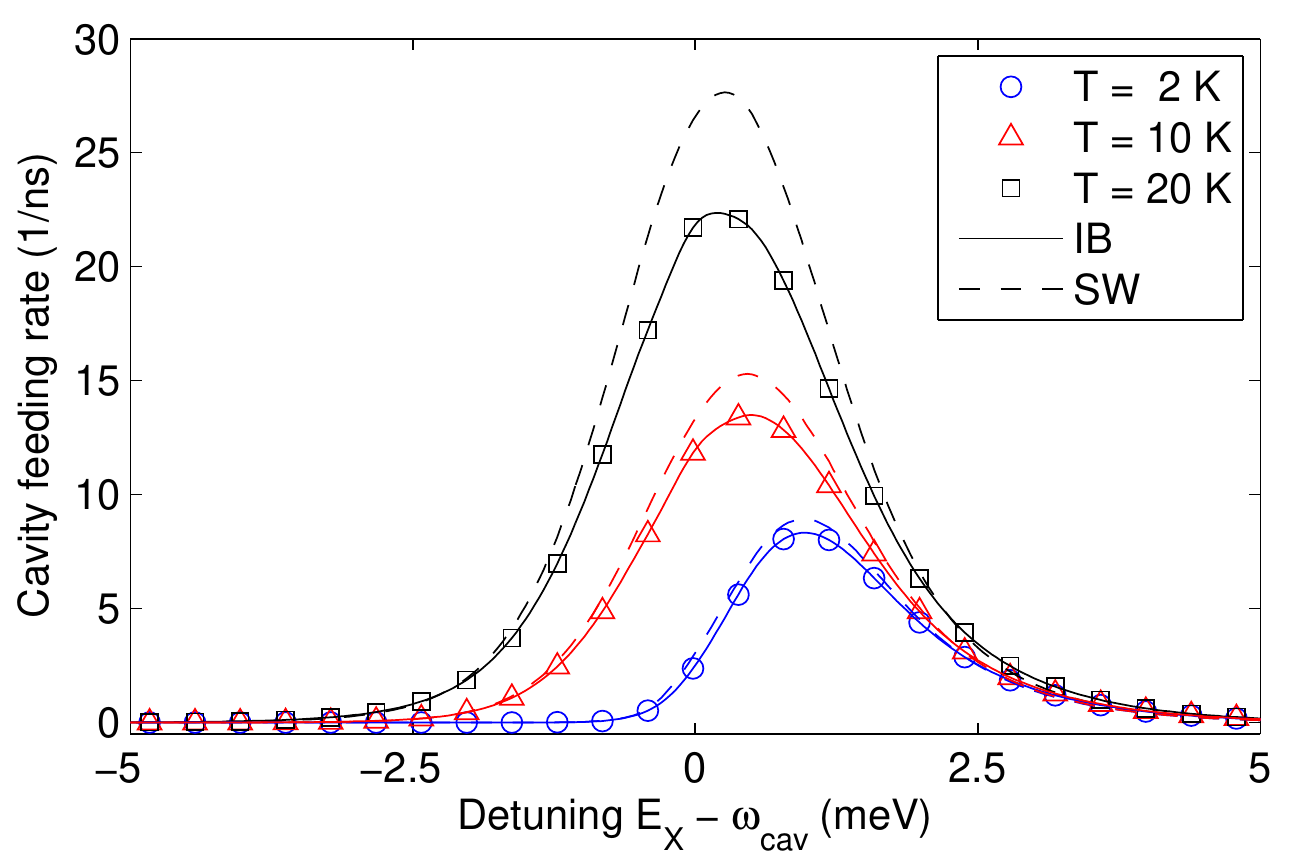}}
\caption{(Color online) Cavity feeding rate for the phonon-assisted transition from the exciton to the cavity. We use the dot and material parameters listed in Table.~\ref{table:param} and an exciton-cavity coupling $g=150\,\mu$eV. The solid and dashed lines report results obtained from Eqs.~\eqref{eq:scattrate.ib} and \eqref{eq:feedingrate} for the independent Boson model (IB) and the Schrieffer-Wolff (SW) perturbation approach, respectively.}
\label{fig:scattrate}
\end{figure}

Figure~\ref{fig:scattrate} shows the cavity feeding rates for the phonon-assisted transition from the exciton to the cavity (and the dot ends up in the groundstate). The solid lines report results from the independent Boson model, Eq.~\eqref{eq:scattrate.ib}, and the dashed lines results of Eq.~\eqref{eq:feedingrate} for the Schrieffer-Wolff perturbation approach. At low temperatures, the feeding rate $\Gamma_{eg}$ as a function of detuning is strongly asymmetric. For positive detuning $\Delta=E_e-E_g-\omega_{\rm cav}$, i.e., when the exciton has a higher energy than the cavity mode, the transition from the exciton to the cavity is accompanied by a {\em phonon emission}.\/ In this regime there is a substantial scattering probability even at low temperatures. With increasing temperature the phonons become thermally populated, as described by the Bose-Einstein factors $\bar n_k$ in the scattering rates of Eqs.~\eqref{eq:scattrate.ib} and Eq.~\eqref{eq:feedingrate}, leading to a noticeable scattering probability also for negative detunings. These processes are  associated with {\em phonon absorption}. At sufficiently high temperatures, say above 20 K, the feeding rate as a function of detuning becomes almost symmetric.

The dependence of the feeding rate on $\Delta$ is governed by the exciton form factor, as given by the integral in Eq.~\eqref{eq:epmatrix}. For the Gaussian carrier wavefunctions, the width of the Fourier-transformed probability distribution is again a Gaussian, with a width given by the inverse of the confinement length $L$. Due to energy conservation in the scattering, the wavenumber $k$, at which the formfactor has to be computed, is determined by $c_\ell k=\Delta$. Thus, a small confinement length $L$ translates to a broad $k$-space distribution of the formfactor, and, in turn, to a broad $\Gamma_{eg}(\Delta)$ distribution. In contrast, for a larger confinement length the feeding rate as a function of detuning becomes narrower. Additionally, the feeding rate is directly proportional to the Purcell enhancement~\cite{purcell:46} $(g/\Delta)^2$ caused by the photon confinement within the microcavity. We emphasize that the scattering rates of both the independent Boson model and the Schrieffer-Wolff perturbation approach directly scale with $g^2$, and do not depend on the quality factor $Q$ of the cavity. However, $Q$ determines the lifetime of the cavity photon, and thus has a decisive impact on the efficiency of phonon-assisted cavity feeding.

From the comparison of the solid and dashed lines in Fig.~\ref{fig:scattrate} we observe that the results of the Schrieffer-Wolff perturbation approach and the independent Boson model are in almost perfect agreement, for temperatures below say 20 K. For elevated temperatures, the Schrieffer-Wolff approach overestimates the feeding rate. We attribute this to the neglect of re-absorption processes of phonons, which are naturally included within the independent Boson model. From now on we will only consider the feeding rates for the independent Boson model, Eq.~\eqref{eq:scattrate.ib}, although our results would look very similar for $\Gamma_{eg}$ computed within the Schrieffer-Wolff perturbation approach.

\subsection{Spontaneous emission lifetime}

\begin{figure}
\includegraphics[width=\columnwidth]{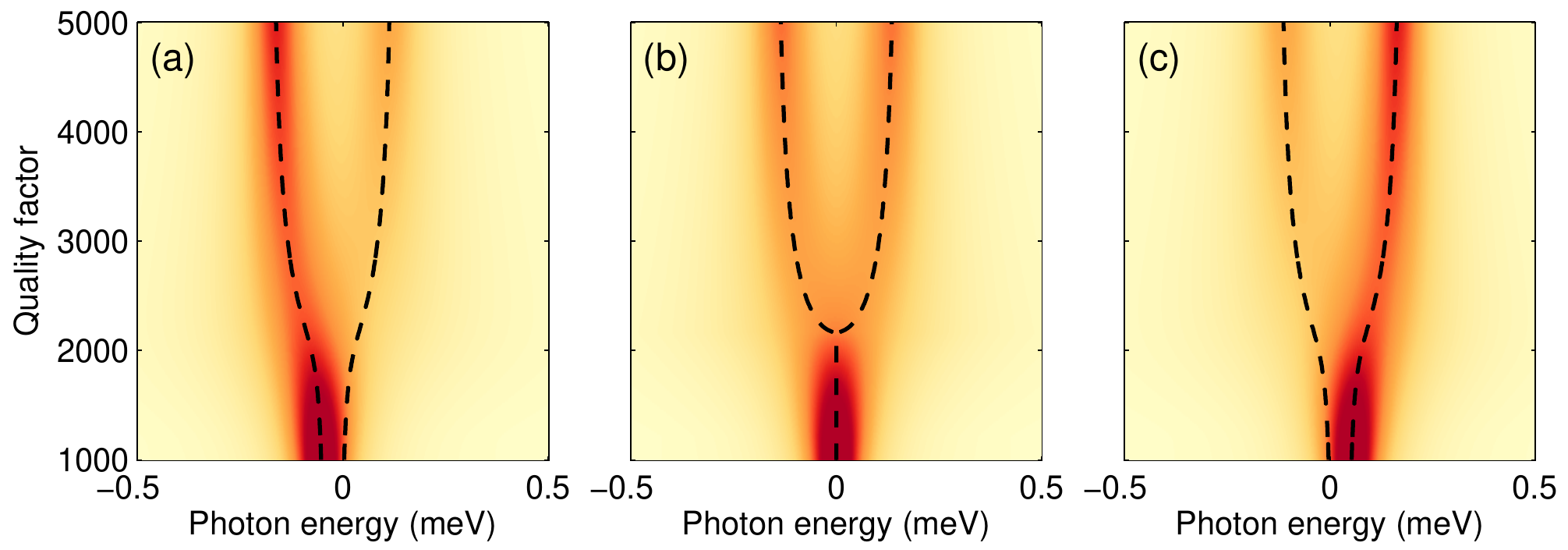}
\caption{(Color online) Spectra for driven dot-cavity system and for different quality factors. We use a level scheme consisting of the quantum dot ground and excited states, and the cavity. The panels report results for a detuning of (a) $\Delta=-50$ $\mu$eV, (b) $\Delta=0$, and (c) $\Delta=50$ $\mu$eV. The temperature is set to $T=10$ K.}
\label{fig:polariton}
\end{figure}

\begin{figure}
\includegraphics[width=\columnwidth]{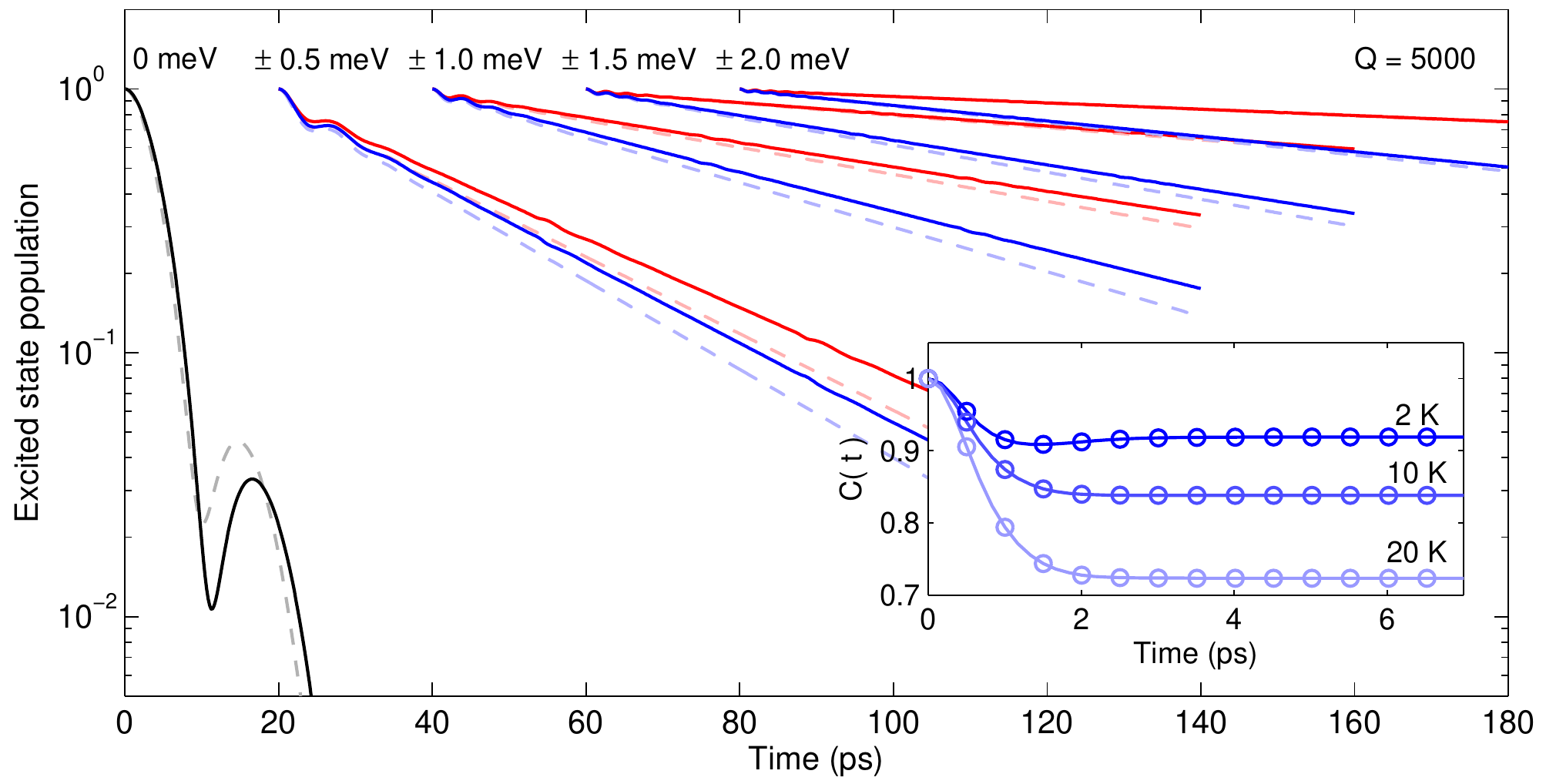}
\caption{(Color online) Exciton decay for a dot embedded in the cavity, and in presence of phonon-assisted transitions. The solid and dashed lines report results of our density matrix and master equation approach, respectively. For clarity, the decay transients for different detunings are offset in time. Throughout the decay for positive (blue) detuning (exciton energy larger than cavity energy $\omega_{\rm cav}$) is faster than for negative (red) detuning. In all simulations we use $Q=5000$ and $T=10$ K. The inset reports results for the correlation function $C(t)$ of the independent Boson model, Eq.~\eqref{eq:correlation}. The solid lines correspond to the analytic expressions, and the symbols show the results of the density matrix approach. Both results are in perfect agreement, thus demonstrating the accuracy of our numerical approach.
}
\label{fig:decay}
\end{figure}

We next consider the situation where the dot is initially in the exciton state, and subsequently decays through either coupling to the leaky cavity mode, phonon-assisted cavity feeding, or radiative decay. From Fig.~\ref{fig:polariton}, which shows luminescence spectra for a driven dot-cavity system, to be discussed further below, we observe that for the chosen dot and cavity parameters the system operates in the strong coupling regime for quality factors above approximately $Q=2000$. Indeed, for $Q=5000$ the decay transients in Fig.~\ref{fig:decay} display at early times population oscillations between exciton and cavity (best visible for zero or small detunings). These oscillations are a clear signature of strong coupling. 

The solid lines report results of our density matrix approach. When comparing the decay for positive and negative detunings $\pm\Delta$, we find that the decay is {\em always faster for positive detunings}\/ than for negative detunings. This is due to the enhancement of phonon emissions in comparison to phonon absorptions, attributed to the different bosonic factors of $\bar n_k+1$ and $\bar n_k$ in the scattering rates. The dashed lines show results of the corresponding master equation simulations. We find nice agreement, with exception of small detunings, where probably higher-order scattering or renormalization processes play some role.

\begin{figure*}
\includegraphics[width=1.65\columnwidth]{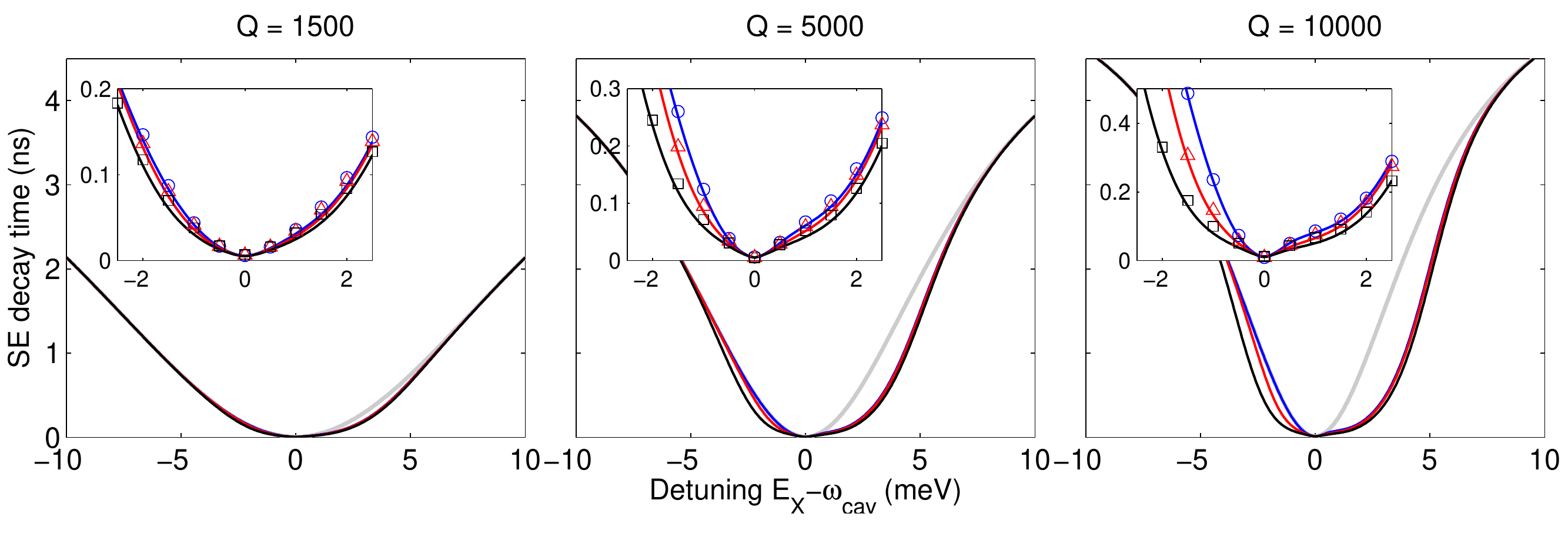}
\caption{(Color online) Spontaneous emission (SE) decay time of initially excited quantum dot for different quality factors and temperatures (same line labeling as in Fig.~\ref{fig:scattrate}). The gray lines report results where phonon scatterings have been artificially neglected. In the insets we compare, for a smaller range of detunings, the density matrix approach (symbols) with the perturbation approach (solid lines), finding perfect agreement throughout.}
\label{fig:sedecay}
\end{figure*}

In the decay of the exciton, the population decays mono-exponentially at times above approximately 20 ps. We can thus assign a lifetime to the excited quantum dot state, which we shall denote as the {\em spontaneous emission}\/ (SE) lifetime. In Fig.~\ref{fig:sedecay} we show the extracted lifetimes $\tau_{SE}(\Delta)$ for a variety of quality factors and temperatures. Quite generally, we find that $\tau_{SE}(\Delta)$ dramatically decreases when exciton and cavity come into resonance. This is attributed to the formation of a polariton with a significant cavity admixture, such that the exciton can decay through the leaky cavity. $\tau_{SE}(\Delta)$ has a broad dip for small quality factors, which significanty narrows for higher $Q$ values.

The gray lines in Fig.~\ref{fig:sedecay} show results of simulations where phonon scatterings have been artificially neglected. When phonon scatterings are included, $\tau_{SE}$ as a function of detuning becomes strongly asymmetric. In particular for the higher quality factors, $\tau_{SE}(\Delta)$ is significantly reduced at positive detunings. This is due to the aforementioned cavity feeding through phonon emission, which is the dominant scattering mechanism at low temperatures. With increasing temperature $\tau_{SE}(\Delta)$ becomes also reduced at negative detunings, as a result of thermally activated phonon absorption. In the insets of the figure we show, for a smaller range of detunings, results obtained within our density matrix approach (symbols) and master equation approach (solid lines). One clearly observes that both approaches give almost identical results, for all the quality factors and temperatures investigated. We thus conclude that the master equation approach with phonon feeding rates, computed from either Eq.~\eqref{eq:feedingrate} or \eqref{eq:scattrate.ib}, provides an accurate description scheme for the problem under consideration. From now on we will only use the master equation approach, described in Sec.~\ref{sec:master}, which provides a simple and versatile simulation tool.

\subsection{Spectrum}

We next discuss the influence of phonon-assisted cavity feeding on the spectral properties of the coupled dot-cavity system. In our simulations we consider the level scheme depicted in Fig.~\ref{fig:scheme}(a), consisting of the dot ground and excited state, and the cavity mode. The system is weakly driven by an external laser pulse, as described by a Lindblad operator which promotes the system from $g$ to $e$. If the pump rate is sufficiently small, in our case much smaller than the radiative decay rate of the quantum dot, only the overall intensities of the spectrum (but not the relative intensities of the upper and lower polariton branches) depend on the pump rate. In our simulations we also include a dephasing rate for the exciton, primarily for visualization purposes to broaden the exciton line. We found that for detunings larger than the polariton linewidths this dephasing has an only minor impact on our results.

\begin{figure}
\includegraphics[width=\columnwidth]{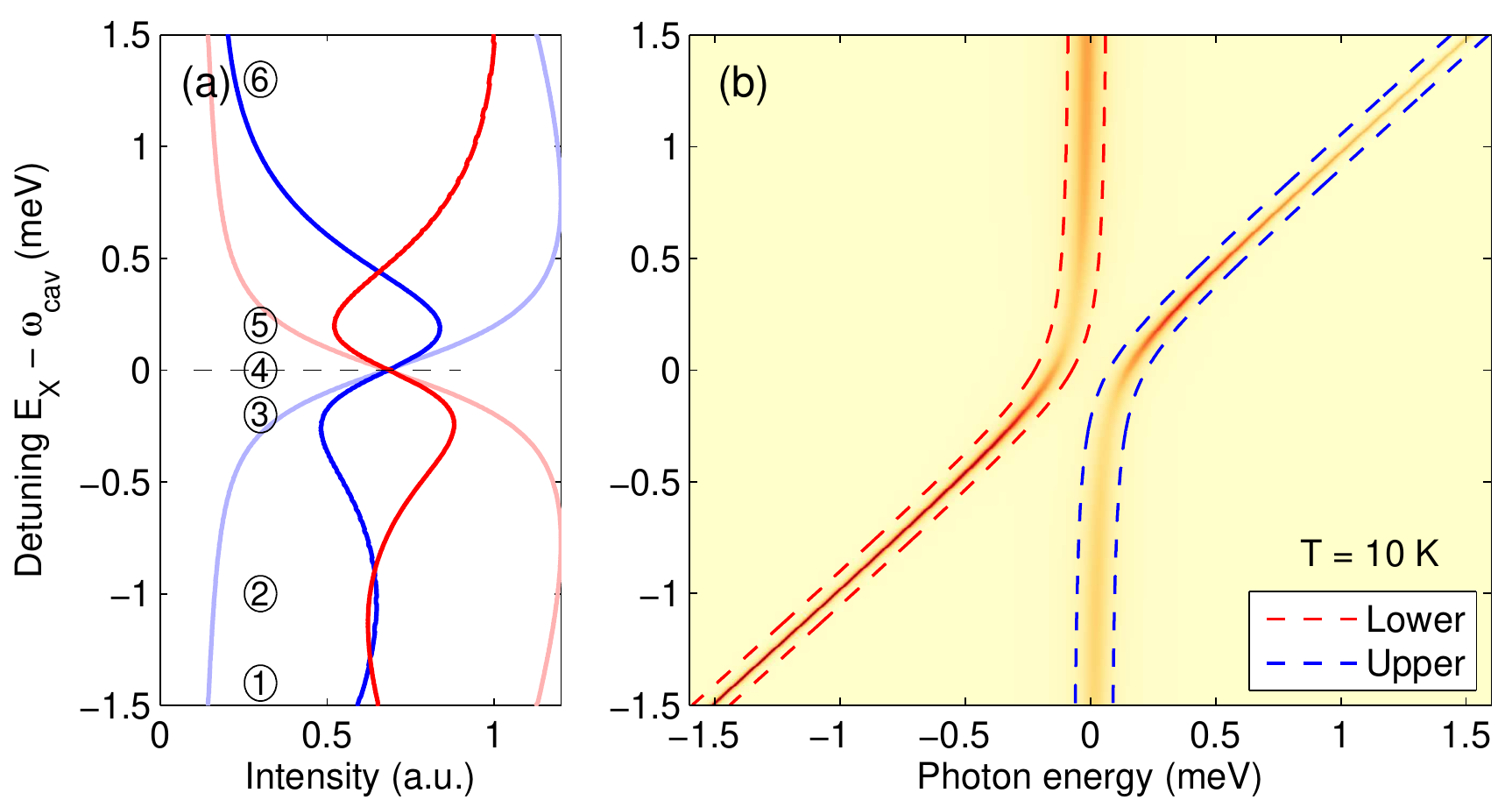}
\caption{(Color online) Spectrum of driven dot-cavity system, as depicted in Fig.~\ref{fig:scheme}. (b) Spectrum as a function of detuning between exciton and cavity. One observes the formation of polariton states, which anticross at zero detuning as a result of strong coupling. In our simulations we use a quality factor of $Q=15000$, a temperature of $T=10$ K, a radiative dot lifetime of $\tau_{\rm rad}=7$ ns, and an exciton dephasing time of 50 ps. (a) Intensity of lower and upper polariton branch. The intensity is integrated over the regions indicated by the dashed lines in panel (b). The bright lines show results of simulations where phonon scatterings have been artificially neglected. The numbers in circles are used for the discussion in the text, and the dashed line indicates the position of zero detuning where the two polariton modes have equal intensity. }
\label{fig:spectrumtwolevel}
\end{figure}

Figure~\ref{fig:spectrumtwolevel}(b) shows the computed spectra for different exciton-cavity detunings. At zero detuning we observe an {\em anticrossing}\/ of the lower and upper polariton branch, which is the signature of strong coupling in the frequency domain.~\cite{walls:95,scully:97,andreani:99,reithmaier:04,hennessy:07} In panel (a) of the figure we plot the intensity of the lower (red) and upper (blue) polariton branch, which is obtained by integrating in the spectra over the spectral region indicated by the dashed lines. 

We observe, in accordance to related work,~\cite{ota:09} a highly non-trivial dependence of the polariton intensity as a function of exciton-cavity detuning. To understand this dependence, we first note that the system is driven through the quantum dot state, and the transition with the higher intensity is the one associated with the quantum dot state, unless there is an efficient feeding of the cavity. Quite generally, the polariton modes can be approximately determined from the eigenvalues and eigenvectors of the effective two-mode hamiltonian~\cite{andreani:99}
\begin{equation}
  H_{\rm eff}=\begin{pmatrix} 
    -i\frac\kappa 2 & g \\ g & \Delta-i\frac\gamma 2 \\ 
  \end{pmatrix}\,,
\end{equation}
where the diagonal terms account for the cavity and excited dot states, including radiative damping, and the off-diagonal terms for the dot-cavity coupling. In the following we discuss for the plait pattern of the polariton intensities, depicted in Fig.~\ref{fig:spectrumtwolevel}(a), the different regions indicated by numbers in circles.
(1) For large negative detunings, the lower polariton branch has a dominant exciton character, and the probability for cavity feeding through phonon absorption (which is needed since the exciton has a lower energy than the cavity) is negligible. Here, the exciton decay becomes enhanced through the cavity, but the system decays at the exciton frequency. In this regime the lower polariton branch has the higher intensity. 
(2) When the detuning is reduced, phonon-assisted feeding becomes more likely, due to the small energy mismatch, and the upper polariton (which has a predominant cavity contribution) emits with a slightly higher intensity. We will show in a moment that this cavity transfer through phonon absorption can be suppressed at even lower temperatures. 
(3) When the detuning is further reduced, the lower polariton branch acquires a larger contribution from the cavity, and can thus decay more efficiently through the leaky cavity. In this regime the lower polariton has a higher intensity. 
(4) At zero detuning the two polariton modes have equal exciton and cavity character, and the intensities of the two polariton branches are equal. 
(5) For positive detuning, the upper (blue) polariton branch has a stronger exciton character. When the detuning is sufficiently small, the upper polariton has a noticeable cavity contribution, and thus emits with higher intensity (analogously to the case of red detuning). 
(6) At larger positive detunings phonon-assisted cavity feeding through phonon emission becomes the dominant mechanism, and the lower (cavity-like) polariton has a much higher intensity. 

From the figure we clearly observe a strong asymmetry between negative and positive detunings, which is attributed to the different probabilities for cavity feeding through phonon absorption or emission. These results are in complete accordance to the above discussion about the spontaneous emission lifetime. Simulations performed by artificially neglecting phonon scatterings (bright lines) only give a Purcell enhancement around zero detuning. Otherwise the results totally differ from those of the simulations including phonon-assisted cavity feeding, thus highlighting the importance of this scattering channel.

\begin{figure}
\includegraphics[width=\columnwidth]{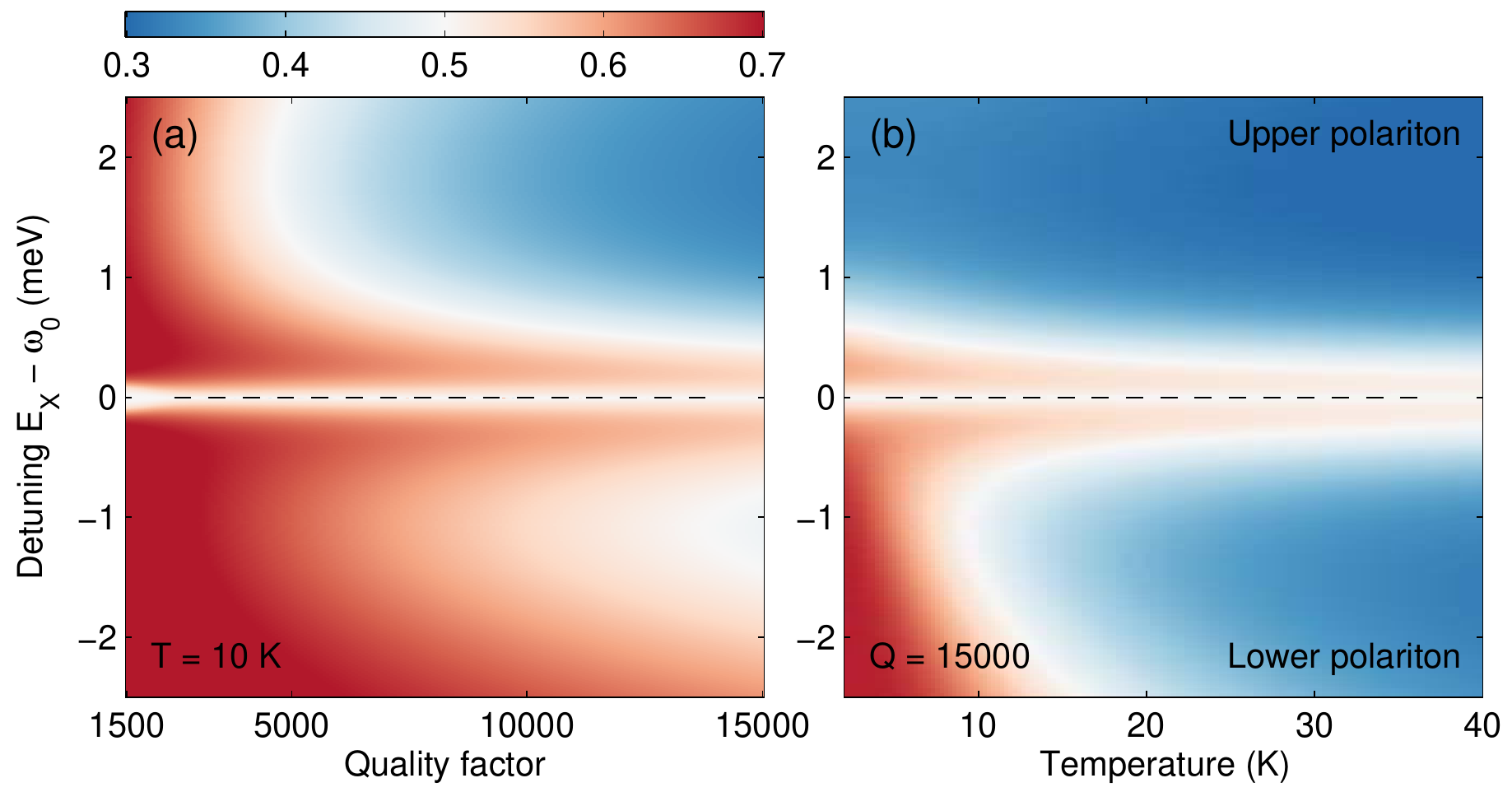}
\caption{(Color online) Relative intensity of quantum-dot-like polariton branch as a function of (a) quality factor $Q$ (with $T=10$ K) and (b) temperature (with $Q=15000$). The dashed lines indicate zero detuning. The relative intensity of the polariton modes is computed by integrating over the area for the lower (red) [upper (blue)] polariton for negative [positive] detunings, as indicated in Fig.~\ref{fig:spectrumtwolevel}(b), and dividing by the sum of the upper and lower polariton intensity. For discussion see text. }
\label{fig:polaritonint}
\end{figure}

In Fig.~\ref{fig:polaritonint} we investigate the polariton intensities for different quality factors and temperatures. Panel (a) reports the influence of the quantum-dot-like polariton (lower polariton branch for negative detunings, and upper branch for positive detunings) intensity on the quality factor (we use $T=10$ K). For small $Q$ values, say below 2500, this branch is strongest for all detunings. For larger $Q$ values the effect of cavity feeding becomes more important, and the intensity significantly drops for positive detunings where the exciton can decay into a cavity photon. We observe that the plait pattern intensity dependence as a function of detuning is most pronounced for high $Q$ values. Here, the cavity photon has the longest lifetime, and accordingly phonon-assisted cavity feeding becomes most efficient. Nevertheless, the influence of $Q$ on the polariton intensity is not overly pronounced in this strong-coupling regime. Panel (b) reports the relative intensity of the lower polariton branch as a function of temperature. The values at 10 K exactly corresponds to the graph of Fig.~\ref{fig:spectrumtwolevel}, which we have discussed above. We observe that cavity feeding through phonon absorption (point 2 of the above discussion) becomes completely suppressed at low temperatures. On the other hand, feeding gains importance at elevated temperatures. At temperature above 30 K the system already predominantly emits at the cavity frequency. 

We note that the plait pattern of the intensity also depends decisively on other feeding channels of the cavity. Such feeding is usually present in experiments due to cavity-assisted relaxation processes of carriers initially excited in the wetting layer,~\cite{kaniber:08,winger:09}. We model it by an additional Lindblad operator that brings the system directly from $g$ to the cavity (see Fig.~\ref{fig:scheme}). Our results (not shown) reveal that the characteristic plait pattern almost disappears when the cavity feeding rate is twice as large as the dot pumping rate. Thus, the polariton intensity dependence could serve as a sensitive measure for additional cavity feeding processes.

\subsection{Biexciton decay and driven biexciton system}

\begin{figure}
\includegraphics[width=\columnwidth]{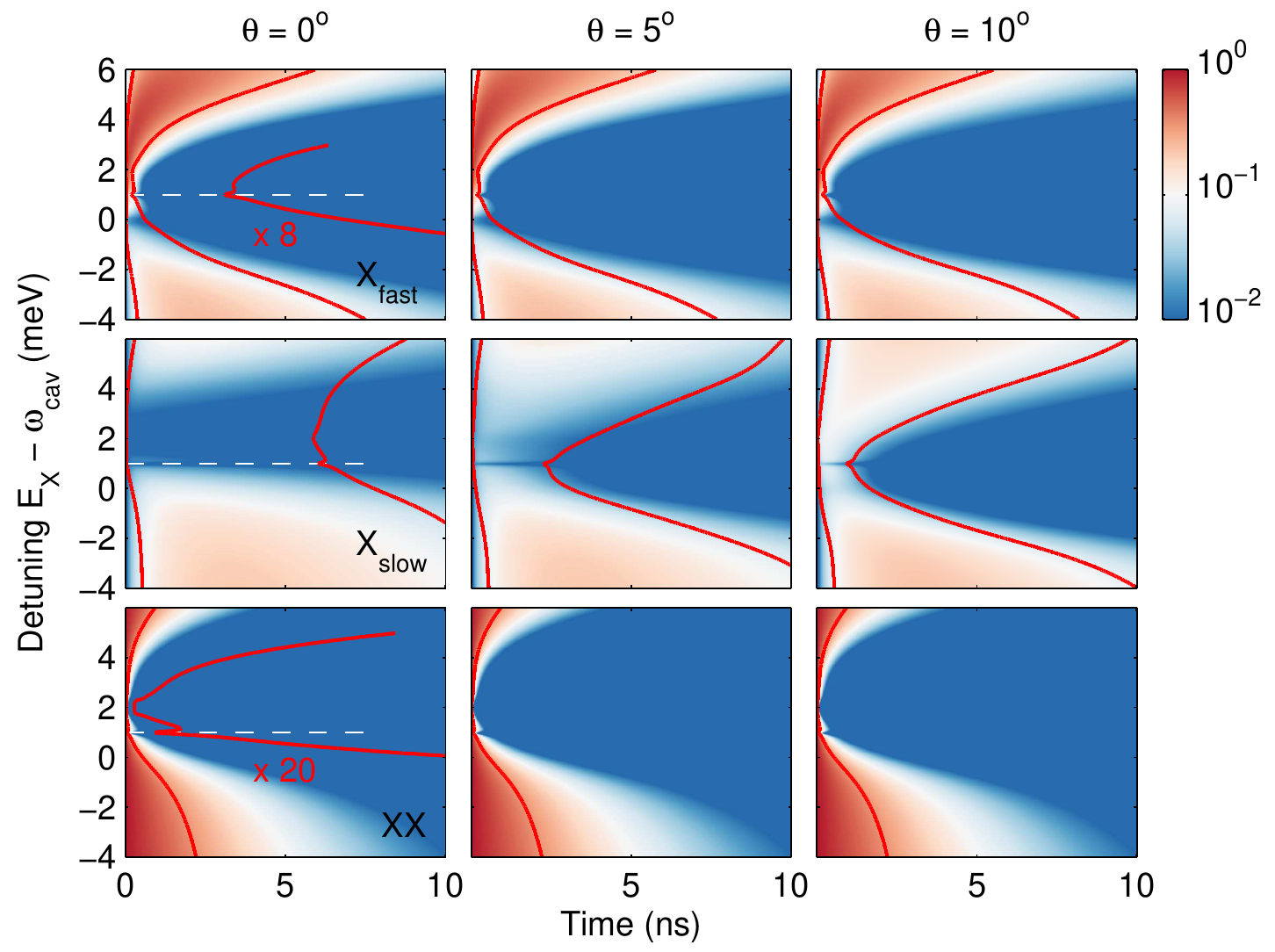}
\caption{(Color online) Time evolution of biexciton and exciton population in the biexciton cascade decay. In the simulations, at time zero only the biexciton is populated, which decays via the exciton states to the ground state. The different rows report the population transients for the exciton $X_{\rm fast}$, that couples with $g\cos\theta$ strongly to the cavity, the exciton $X_{\rm slow}$, that couples with $g\sin\theta$ weakly to the cavity, and the biexciton $X\!X$. The different columns report results for different mixing angles $\theta$ [see Fig.~\ref{fig:scheme}(b)]. In our simulations we use $Q=15000$ and $T=10$ K. The contour lines indicate where the population has decayed to 37\% of its maximum value. For the fast exciton and the biexciton decay these contour lines are also magnified to make the features at early times better visible. The dashed lines indicate the position where the biexciton is in resonance with two cavity photons.}
\label{fig:bidecay}
\end{figure}

In the following we consider a more complete level scheme for the quantum dot, depicted in Fig.~\ref{fig:scheme}(b), consisting of the dot groundstate, two exciton states $x$ and $y$ with orthogonal, linear polarization, and the biexciton state $X\!X$. We also include states where, in addition, one or two photons are present in the cavity. We will primarily investigate how our conclusions for the two-level system, discussed above, become modified when considering this more complete dot description.

From the beginning we introduce a few simplifications. First, we neither consider (spin forbidden) dark states, which might play some role for small detunings or for the driven biexciton system,~\cite{winger:08} nor charged exciton states, which seem to be ubiquitous for quantum dots embedded in nanocavities.~\cite{hennessy:07,kaniber:08} Also the finestructure splitting between the two exciton states is ignored. In principle, such splitting could be easily included in our simulations, but we do not expect that the resulting modifications are of relevance for the relatively large detunings (up to several meV) considered here. In our simulations we assume that both exciton states $x$ and $y$ can couple to the cavity, and introduce a mixing angle $\theta$ that determines the relative coupling strengths of $g\cos\theta$ and $g\sin\theta$, respectively. We will only be interested in small mixing angles, and will consequently denote the excitons with the fast ($g\cos\theta$) and slow ($g\sin\theta$) decay characteristics as $X_{\rm fast}$ and $X_{\rm slow}$, respectively. 

\begin{figure}
\includegraphics[width=\columnwidth]{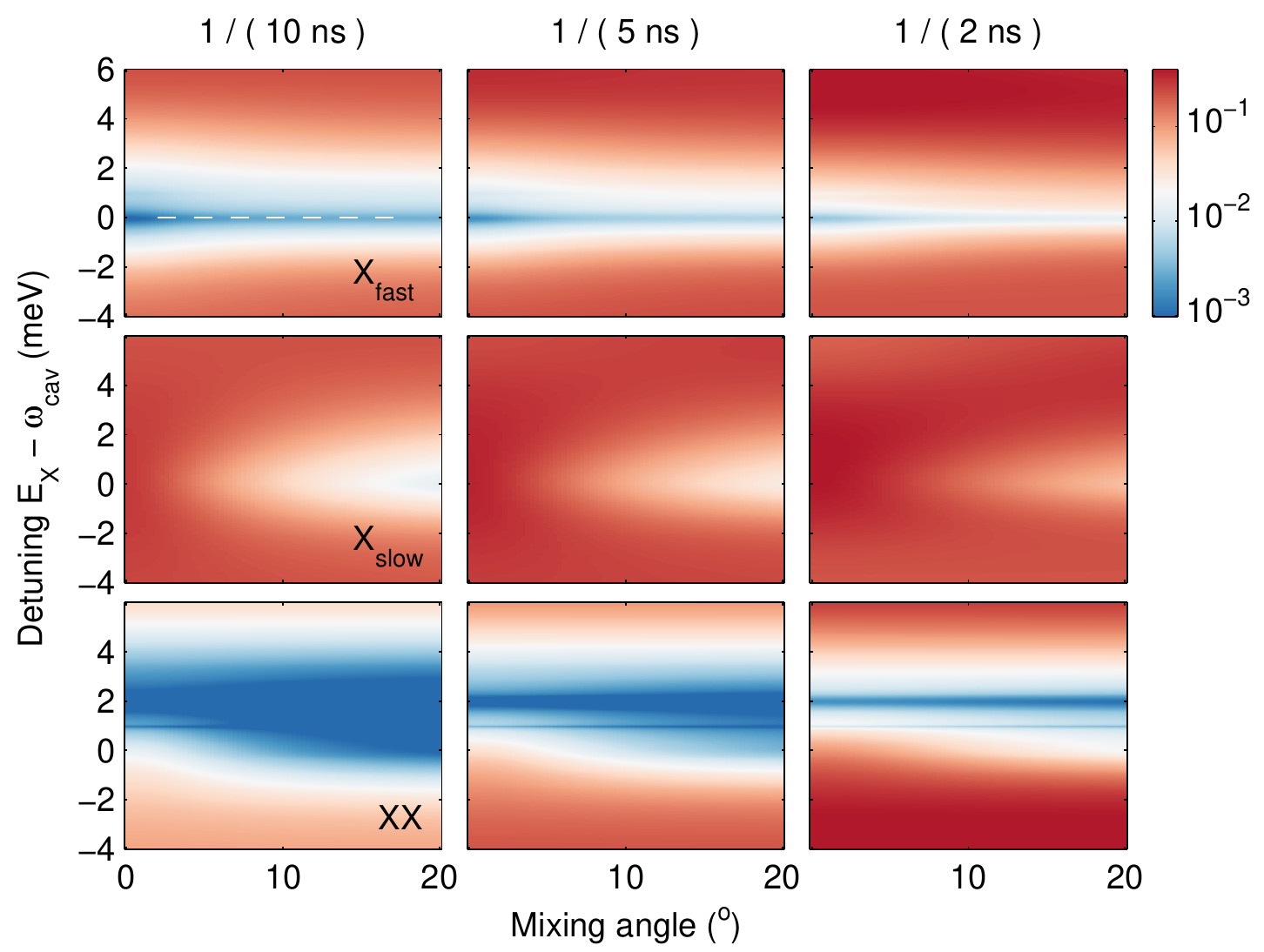}
\caption{(Color online) Steady state exciton and biexciton populations of a driven biexciton system. We use $Q=15000$ and $T=10$ K. The labeling of the exciton and biexciton states is identical to Fig.~\ref{fig:bidecay}. The different columns correspond to different pumping rates, indicated on top of the panels. }
\label{fig:population}
\end{figure}

Figure~\ref{fig:bidecay} shows simulation results of a dot that is initially populated with a biexciton, and subsequently decays through radiative couplings and phonon-assisted transitions to the cavity. No pumping is considered here. The situation under investigation approximately corresponds to experiments where at time zero electron-hole pairs are excited in the wetting layer, or in excited dot states, and consecutively relax on a short timescale to the biexciton state of lowest energy. Let us first consider a zero mixing angle (left column). In accordance to the spontaneous emission lifetimes of the exciton, shown in Fig.~\ref{fig:sedecay}, we observe for the biexciton (lower left panel) an asymmetric detuning dependence for the population decay, which, however, is less pronounced than for the exciton. The fastest decay is now at a detuning corresponding to the biexciton binding energy $\Delta_b$ of 2 meV, where the biexciton to exciton decay is in resonance with the cavity. 

For our quantum dot model, the phonon-assisted cavity feeding rates for the biexciton are the same as for the exciton. According to the discussion of Sec.~\ref{sec:theory}, we have to use in the scattering rates the difference  $\lambda_{ee,k}-\lambda_{gg,k}$ of biexciton and exciton phonon coupling constants. If we neglect possible modifications of the carrier wavefunctions in the biexciton state, which is certainly a good approximation that will be adopted in this work, the scattering rates for phonon-assisted biexciton and exciton decay are the same, apart from the inclusion of the biexciton binding $\Delta_b$ in the detunings.

A striking feature of the biexciton decay depicted in Fig.~\ref{fig:bidecay} is the sharp feature around 1 meV, best visible from the contour line that indicates where the population has dropped to 37\% of its initial value (see closeup where the time scale has been reduced by a factor of 50). The detuning of 1 meV corresponds to the two photon resonance condition
\begin{equation}
  2E_x-\Delta_b=2\omega_{\rm cav}\,,
\end{equation}
where the biexciton can decay by emitting two cavity photons simultaneously.~\cite{hohenester.prb:02,stufler:06b} Here $E_x$ is the exciton energy. Because the transition involves a tunneling-like transition, through the off-resonant exciton state, the detuning width where this process is efficient is extremely narrow. 

Through the biexciton decay, the exciton states become populated and consecutively further decay to the groundstate. Let us first concentrate on the fast exciton component $X_{\rm fast}$. We observe for the population decay a strongly asymmetric detuning dependence, in accordance to our previous discussion for the two-level dot model. However, the minimum is now shifted to larger detunings. This is due to the fact that the biexciton decays fast, and correspondingly the exciton population builds of quickly. Thus, the exciton decay reflects the combined effects of population buildup, through the biexciton decay, and exciton decay. Incidentally, the fastest exciton decay is at the detuning where the biexciton is in resonance with two cavity photons, and also the exciton decay is fast as a result of the Purcell enhancement. We note that for this detuning also the population that is channeled through the exciton is smallest, since the biexciton can bypass the excitons in the direct two-photon decay.

We next investigate the decay of the slow exciton component $X_{\rm slow}$. For zero mixing angle, $\theta=0$, exciton feeding (through the biexciton) and decay are not enhanced by the cavity. We observe in Fig.~\ref{fig:bidecay} that for sufficiently large negative detunings, where the biexciton decays without a significant Purcell enhancement, both fast and slow exciton components are populated equally. $X_{\rm slow}$ decays with the radiative decay time $\tau_{\rm rad}$, in contrast to the fast component, whose decay is enhanced through cavity coupling. For increased detunings, the biexciton decay becomes enhanced through phonon feeding and the Purcell effect, and decays primarily via the $X_{\rm fast}$ channel. Thus, for $\theta=0$ the decay of the slow exciton component directly mirrors the decay characteristics of the biexciton. In presence of a non-zero mixing angle $X_{\rm slow}$ acquires a small cavity admixture, through the coupling $g\sin\theta$, and its feeding and decay become enhanced. Indeed, we observe a pronounced lifetime decrease for small cavity detunings, with the minimum being at the two-photon resonance. 


Finally, in Fig.~\ref{fig:population} we analyze the steady state populations of the driven biexciton system depicted in Fig.~\ref{fig:scheme}. We chose pumping rates, indicated on top of the panels, which are smaller or larger than the radiative decay rates of the isolated exciton states. For essentially all pumping rates we observe a significant population of the biexciton state. This is due to the fact that the $X_{\rm slow}$ population is usually large, due to the long lifetime originating from the weak cavity coupling. Thus, there exists an efficient population channel of the biexciton via the intermediate $X_{\rm slow}$ state. The steady state populations of all states reflect the respective decay characteristics. The populations are large when the lifetime is long, and small when the lifetime is short. As expected, with increasing pumping rate the steady state populations increase.

\subsection{Biexciton spectra}

\begin{figure}
\includegraphics[width=\columnwidth]{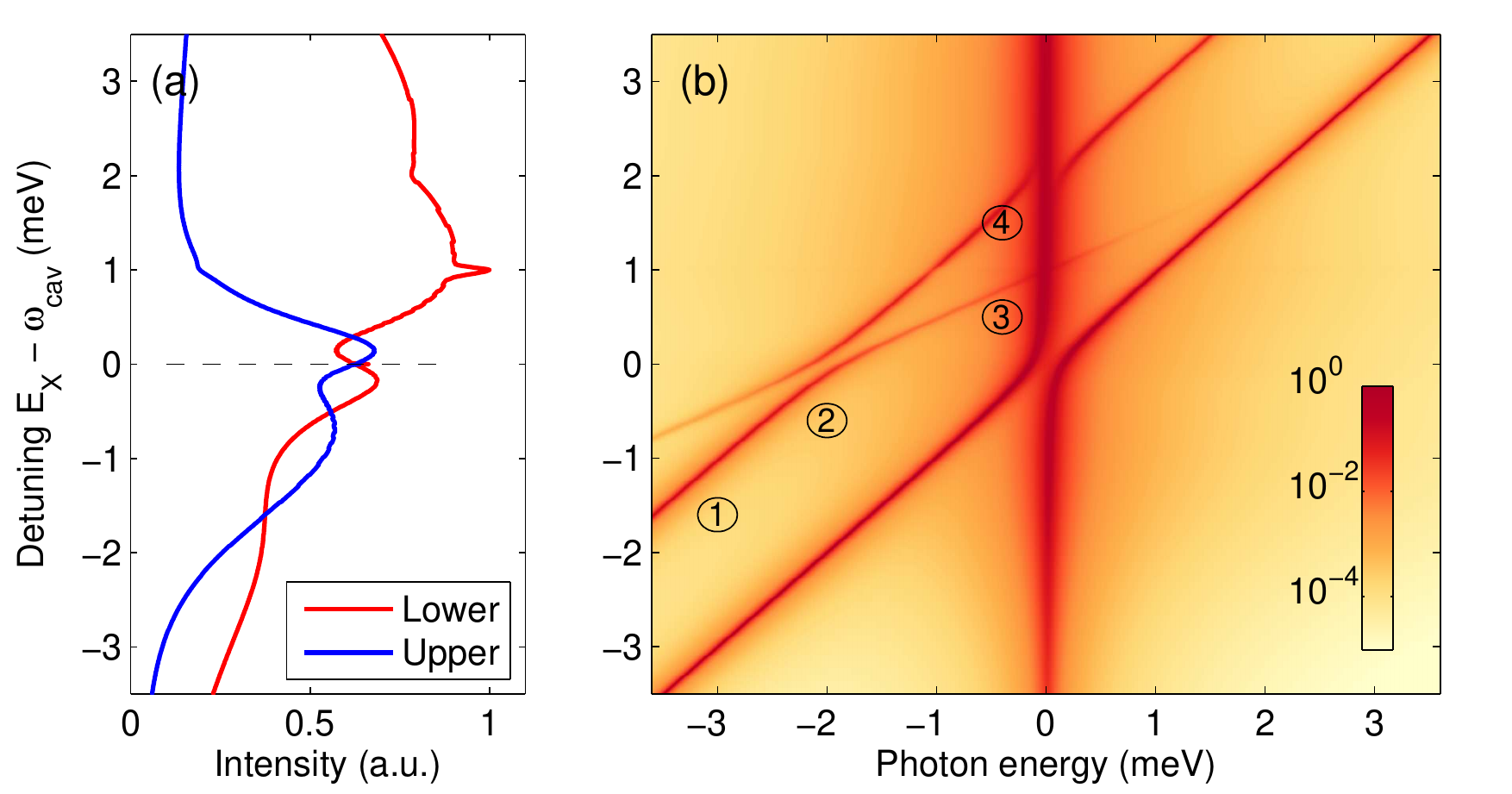}
\caption{(Color online) Simulated spectra for driven biexciton system. We use  $Q=15000$, $T=10$ K, a mixing angle of $\theta=5^\circ$, and a pump rate of $1/(5\, {\rm ns})$. To make the biexcitonic features better visible, we use a logarithmic color scale. The intensities of the polariton modes are computed in accordance to the prescription given in the caption of Fig.~\ref{fig:spectrumtwolevel}. The numbers in circles are used in the discussion in the text.}
\label{fig:spectrum}
\end{figure}

We finally compute luminescence spectra for a driven biexciton system. Figure~\ref{fig:spectrum} shows the spectra as a function of detuning. We use a logarithmic color scale to make the weak biexciton features better visible. In the luminescence spectra we observe, in addition to the lower and upper polariton branches of the coupled exciton-cavity system, peaks that are shifted by approximately 2 meV to the red, as indicated in the figure by (1). These features are associated with the biexciton to exciton decay. (2) The anticrossing at zero detuning is attributed to the biexciton decay to the strongly coupled upper and lower exciton-polaritons.~\cite{winger:08} At point (3) we observe a crossing with the lower exciton-polariton branch. As the two branches are associated with the cavity-like polaritons, this crossing point precisely indicates the point where the biexciton is in resonance with the two-photon transition. (4) Finally, when the biexciton to exciton transition is in resonance with the cavity mode, we observe an anticrossing of the polaritons associated with this transition. At the same spectral position there is additonally the emission of the cavity-like polariton of the coupled exciton-cavity system.

The intensity of the lower (red) polariton branch, shown in panel (a), exhibits a small bump at the point of two-photon resonance. When comparing the results of Fig.~\ref{fig:spectrum}(a) with those of the two-level system, Fig.~\ref{fig:spectrumtwolevel}(a), we find that the relative intensities are very similar, and only the detuning dependence of the absolute intensities differ. This is due to the more complicated excitation dynamics in case of biexciton pumping. Nevertheless, the main conclusions for the dot-cavity system regarding quality factor and temperature prevail.


\section{Summary and conclusions}\label{sec:summary}

To summarize, we have investigated phonon-assisted transitions for a coupled quantum dot-cavity system, where the exciton decays into the cavity and the energy mismatch is compensated through phonon emission or absorption. Such processes play an important role for energy detunings of the order of a few meV, and for dot-cavity systems operating in the strong coupling regime. The basic mechanism underlying phonon-assisted cavity feeding can be encapsulated in the effective Hamiltonian
\begin{equation}\label{eq:conclusion}
  H'=\frac g\Delta
  \left(\sigma_{eg}a-\sigma_{ge}a^\dagger\right)
  \sum_k\lambda_{ee,k}\left(b_k^{\phantom\dagger}+b_k^\dagger \right)\,,
\end{equation}
which has been derived within the framework of a Schrieffer-Wolff transformation [Eq.~\eqref{eq:phonariton}]. It describes a combined phonon and cavity effect, where $\sigma_{ge}a^\dagger$ accounts for the exciton decay and the creation of a cavity photon, and the phonon field operators $b_k^\dagger$ and $b_k^{\phantom\dagger}$ for the accompanying emission or absorption of a phonon. Although such processes can in principle also occur for quantum dots outside a microcavity, they become strongly enhanced within a cavity, because of the Purcell enhancement of $g/\Delta$.

In addition to the Schrieffer-Wolff perturbation approach, we have derived a similar expression for the independent Boson model [Eq.~\eqref{eq:scattrate.ib}], where the phonon dynamics has been accounted for in an exact manner. We have also analyzed the problem within a numerical density matrix approach, without employing any perturbation procedure. For realistic dot and cavity parameters, all approaches give more or less the same results, at least for temperatures below 20 K. 

We have investigated the role of phonon-assisted cavity feeding in the exciton and biexciton decay, and have found a strong detuning asymmetry in agreement with experiment.~\cite{hohenester.prb:09} This has been attributed to the different importance of phonon emissions and absorptions. For the spectra of a driven dot system, we have found an intriguing plait pattern for the intensities of the upper and lower polariton branches.~\cite{ota:09} We have shown that these features prevail for more realistic dot level schemes. For the biexciton cascade decay, we have analyzed the different decay channels via the intermediate exciton states, and have found a fast and slow decay component, in agreement with experiment.~\cite{hohenester.prb:09} For a driven biexciton state, an efficient population channel of the biexciton via one of the exciton states (weakly coupled to the cavity) has been identified.

Phonon-assisted cavity feeding plays a dominant role in cavity-QED experiments whenever the dot and cavity are detuned by only a few meV, and might be of importance for quantum dot-based lasers,~\cite{nomura:06} single or entangled photon sources,\cite{michler:00,stace:03,stevenson:06,akopian:06} or in quantum information science.~\cite{mabuchi:02}

\begin{acknowledgements}

I am grateful to Arne Laucht, Jonathan Finley, Martin Winger, and Atac Imamoglu for most helpful discussions.

\end{acknowledgements}

\begin{appendix}

\section{Density matrix approach}\label{sec:cumulant}

In this appendix we give some details of our density matrix approach. For each correlation function $\langle\!\langle A\rangle\!\rangle$ we derive the dynamic equation from the Heisenberg equation of motion
\begin{equation}
  i\frac d{dt}\bigl< A\bigr>=\bigl<
  AH_{\rm eff}^\dagger-H_{\rm eff}^{\phantom\dagger}A\bigr>\
  +\bigl<[A,\sigma_{ee}]B_{ee}\bigr>\,,
\end{equation}
and subtract the contributions from the lower-order density matrices. In the following we only consider the non-trivial terms according to phonon coupling. The contributions due to $H_{\rm eff}$ can be added in a straightforward fashion.

As for the density matrix of the two-level system we obtain
\begin{subequations}
\begin{eqnarray}
  i\dot\rho_{ge}&=&\phantom{-}\Omega\rho_{ge}+\Omega_{ge}\\
  i\dot\rho_{eg}&=&-\Omega\rho_{eg}-\Omega_{eg}\,,
\end{eqnarray}
\end{subequations}
together with $\dot\rho_{gg}=\dot\rho_{ee}=0$. Here we have introduced the abbreviations
\begin{subequations}
\begin{eqnarray}
  \Omega &=& \sum_k\lambda_{ee,k}\bigl(s_k^{\phantom *}+s_k^*\bigr)\\
  \Omega_{eg} &=& \sum_k\lambda_{ee,k}\bigl(s_{eg,k}^{\phantom *}+s_{ge,k}^*\bigr)\,
\end{eqnarray}
\end{subequations}
for the energy renormalization $\Omega$, as well as the source terms $\Omega_{eg}$ and $\Omega_{ge}^{\phantom*}=\Omega_{eg}^*$, which describe the coupling to the phonon-assisted density matrices. For the time evolution of the phonon density matrices we get
\begin{subequations}
\begin{eqnarray}
  i\dot s_k &=&\lambda_{ee,k}\rho_{ee}\\
  i\dot s_{kk'} &=&\lambda_{ee,k}s_{ee,k'}+\lambda_{ee,k'}s_{ee,k}\\
  i\dot n_{kk'} &=&\lambda_{ee,k}s_{ee,k'}^*-\lambda_{ee,k'}s_{ee,k}\,,
\end{eqnarray}
\end{subequations}
and for that of the phonon-assisted density matrix
\begin{eqnarray}
  i\dot s_{ge,k} &=& \phantom-\Omega s_{ge,k}+\lambda_{ee,k}
  (1-\rho_{ee})\rho_{ge}+\Omega_k\rho_{ge}+\Omega_{ge,k}\nonumber\\
  i\dot s_{eg,k} &=& -\Omega s_{eg,k}-\lambda_{ee,k}\phantom{(1-}\,\,\rho_{ee}\phantom{)}
  \rho_{eg}-\Omega_k\rho_{eg}-\Omega_{eg,k}\nonumber\\ 
  i\dot s_{ee,k} &=& \lambda_{ee,k}(1-\rho_{ee})\rho_{ee}\nonumber\\
  i\dot s_{gg,k} &=& 0\,.
\end{eqnarray}
We have introduced for the coupling to the higher-order phonon correlations the abbreviations
\begin{subequations}
\begin{eqnarray}
  \Omega_k &=& \sum_{k'}\lambda_{ee,k'}\bigl(s_{kk'}+n_{kk'}\bigr)\\
  \Omega_{eg,k} &=& \sum_{k'}\lambda_{ee,k'}\bigl(s_{eg,kk'}+n_{eg,kk'}\bigr)\,.
\end{eqnarray}
\end{subequations}
For the two-phonon-assisted density matrices we finally obtain 
\begin{subequations}
\begin{eqnarray}
  i\dot s_{ge,kk'} &=& \Omega s_{ge,kk'}+\Omega_k s_{ge,k'}+\Omega_{k'}s_{ge,k}
  \nonumber\\
  &+&\lambda_{ee,k\phantom'}\bigl((1-\rho_{ee})s_{ge,k'}-\rho_{ge}s_{ee,k'}\bigr)\nonumber\\
  &+&\lambda_{ee,k'}\bigl((1-\rho_{ee})s_{ge,k\phantom'}-\rho_{ge}s_{ee,k\phantom'}\bigr)\\
  i\dot s_{eg,kk'} &=& -\Omega s_{eg,kk'}-\Omega_k s_{eg,k'}-\Omega_{k'}s_{eg,k}
  \nonumber\\
  &-&\lambda_{ee,k\phantom'}\bigl(\rho_{ee}s_{eg,k'}+\rho_{eg}s_{ee,k'}\bigr)\nonumber\\
  &-&\lambda_{ee,k'}\bigl(\rho_{ee}s_{eg,k\phantom'}+\rho_{eg}s_{ee,k\phantom'}\bigr)\\
  i\dot s_{ee,kk'} &=& (1-2\rho_{ee})(\lambda_{ee,k}s_{ee,k'}+\lambda_{ee,k'}s_{ee,k})\quad
\end{eqnarray}
\end{subequations}
and
\begin{subequations}
\begin{eqnarray}
  i\dot n_{ge,kk'} &=& \Omega n_{ge,kk'}+\Omega_k s_{eg,k'}^*+\Omega_{k'}^*s_{ge,k}
  \nonumber\\
  &+&\lambda_{ee,k\phantom'}\bigl((1-\rho_{ee})s_{eg,k'}^*-\rho_{ge}s_{ee,k'}^*\bigr)\nonumber\\
  &+&\lambda_{ee,k'}\bigl(\phantom{(1-}\,\,\rho_{ee}\phantom{)}s_{ge,k\phantom'}+\rho_{ge}s_{ee,k\phantom'}\bigr)\\
  i\dot n_{eg,kk'} &=& -\Omega n_{eg,kk'}-\Omega_k s_{ge,k'}^*-\Omega_{k'}^*s_{eg,k}
  \nonumber\\
  &-&\lambda_{ee,k\phantom'}\bigl(\phantom{(1-}\,\,\rho_{ee}\phantom{)}s_{ge,k'}^*+\rho_{eg}s_{ee,k'}^*\bigr)\nonumber\\
  &-&\lambda_{ee,k'}\bigl((1-\rho_{ee})s_{eg,k\phantom'}-\rho_{eg}s_{ee,k\phantom'}\bigr)\\
  i\dot n_{ee,kk'} &=& (1-2\rho_{ee})(\lambda_{ee,k}s_{ee,k'}^*-\lambda_{ee,k'}s_{ee,k})\,,\qquad
\end{eqnarray}
\end{subequations}
together with $\dot s_{gg,kk'}=\dot n_{gg,kk'}=0$. This provides us with a closed set of equations. In our computational approach, we introduce a surrogate hamiltonian, with a finite number of representative phonon modes (typically a few hundred),~\cite{hohenester.jpb:07} and solve the equations of motion through direct numerical integration.

\end{appendix}


\begin{thebibliography}{57}
\bibitem[{rem()}]{remark.correlation}
\bibinfo{note}{Note that the correlation of Eq.~\eqref{eq:correlation}, which
  describes pure dephasing, is the complex conjugate of the correlation
  function appearing in Eq.~\eqref{eq:scattrate.ib0}. This is due to the fact
  that in the dephasing description of optically excited quantum dots one
  assumes a transition from the ground to the excited state, in contrast to
  phonon-assisted cavity feeding, where the excited state decays into the
  ground state.}
\end{thebibliography}
\end{document}